# Laser Laboratory Beam Alignment Skills: Course Package

Glen D. Gillen

California Polytechnic State University, San Luis Obispo

Abstract

A series of tutorials, assessments, and instructor guides are presented as a complete package for an upper-level undergraduate, or lower-level graduate, laboratory-based course, or extended new-student seminar. The purpose of this package is to teach the students essential skills beneficial for working in an experimental optical laboratory by introducing them to fundamental laboratory skills, and advanced optical alignment techniques. It is assumed that the students do not have any prior optical laboratory training or experience and the tutorials are written using detailed step-by-step instructions for the students to follow independently without the need for continual instructor guidance. The tutorials are intended to establish a common fundamental knowledge base and set of optical alignment skills within the group of students taking the course. Two examples of the target student groups for this package are: (1) an upper-level undergraduate optics course with a laboratory component (i.e., an undergraduate institution similar to Cal Poly), or (2) a cohort of new graduate students to an experimental optical program whose undergraduate backgrounds can span anything from no experimental optical experience to extensive experimental optical experience (i.e., an experimental R1 doctoral university similar to the Institut für Angewandte Physik at the Universität Bonn). The accompanying assessments and instructor guides for each tutorial are provided in this package for the convenience of the course instructor(s) to use in the course as-is, or to use for their own reference of what skills to assess and how to tell if the student did the tasks correctly.

# Table of Contents

# Tutorial 1: Laser output along a table bolt-hole line

**Tutorial Goals**

The goals of this tutorial are:

- to understand fundamental laser safety procedures and protocols
- to position the laser such that the output beam *exactly* follows above a bolt-hole line of the optics table
- to learn how to properly secure the laser and bolt it to the optics table

**Laser Safety**

As this is the beginning of the laser skills tutorials, some basic laser safety training is necessary.

- **Always assume a laser is dangerous** and treat with caution and respect.  There are different classifications of lasers depending upon the laser's power and ability to cause damage to the human eye and skin.  Even a "safe" class 1 or 2 laser can cause temporary or permanent retinal damage under the right conditions.  Class 4 lasers can cause permanent retinal damage in a fraction of a second.  **In other words, a momentary mistake can lead to permanent blindness.**
- **Never look into the output of a laser.  Even if you think it is off, or not working.** This may seem obvious, but habitual behavior is hard to change.  For example, if something does not seem to be working, or turning on, it is human nature to look directly at it to figure out why.  A common beginner's mistake in an optical lab occurs when the laser is not turning on and someone instinctually either bends down to look at the output of the laser or picks the laser up and points it towards their face.

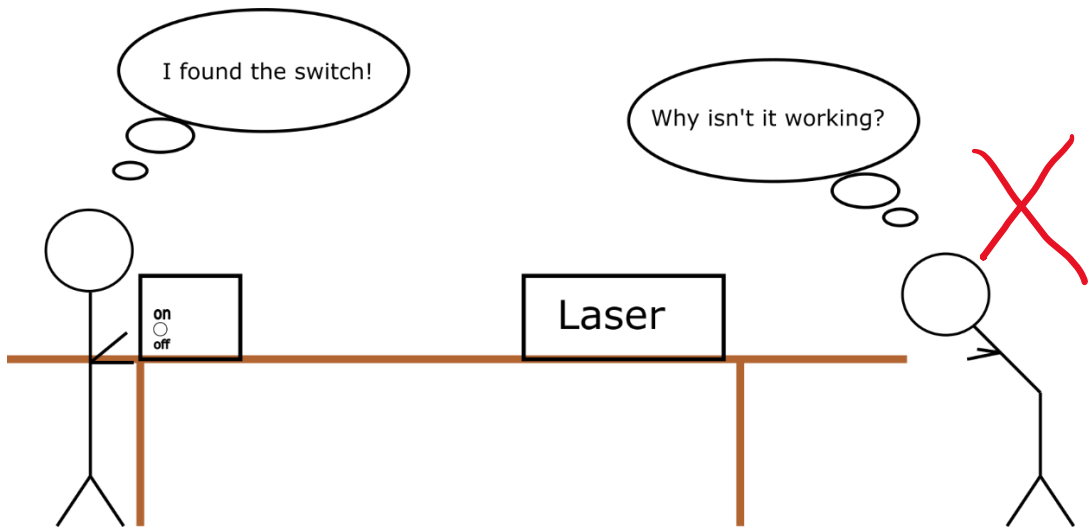

- **Always use your hand, or a piece of paper, near the output of the laser to observe or locate the beam.**
- **Reflective surfaces caution:** avoid placing any other smooth reflective surfaces in or near your working area with the laser beam; i.e., screw drivers, ball drivers, watches, rings with flat surfaces, etc. If you wear a watch, it is good optics lab protocol to take your watch off if your hands are going to be anywhere near the laser beam.
- **Never bend down, or sit, so that your eye level is near the beam level of the table.** Always stand such that you are looking down at the laser beam, beam path and optical components.

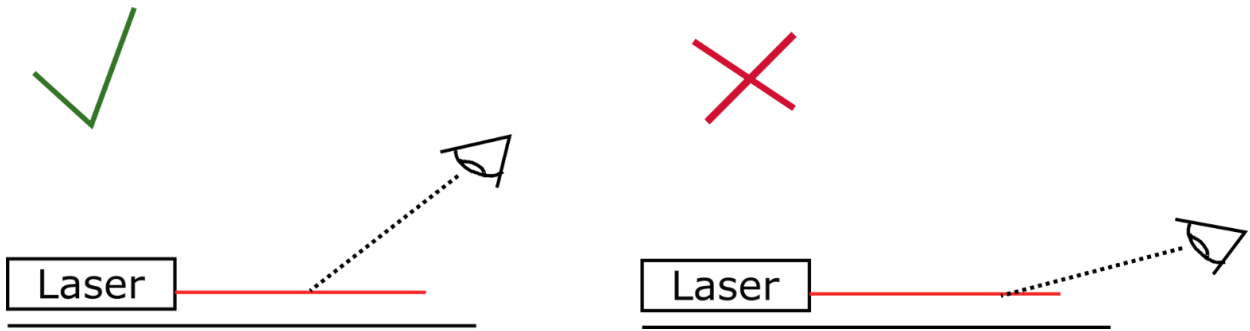

**Turning on the Laser**

- Place the laser on the optics table and place a beam block in front of the laser output.

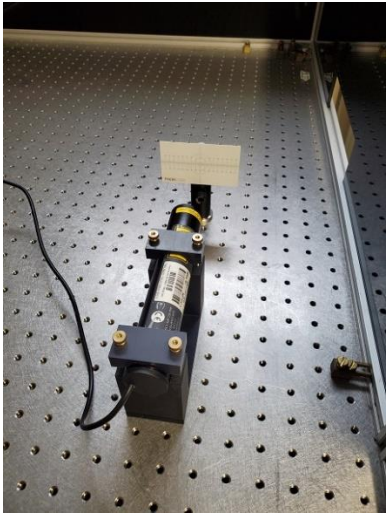

- If the laser has a shutter, set the shutter to the "closed" position.
- Orient the laser housing such that the laser is generally pointing along a chosen bolt-hole line of the optics table.
- Provide power to the laser, turn it on, and if equipped open the shutter. A dot of laser beam should appear on the beam block.

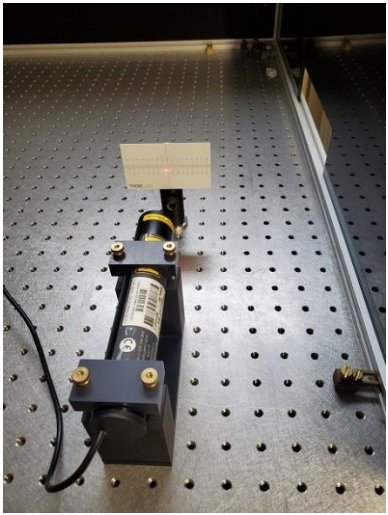

- *Make sure that the beam always terminates on the beam block.*

**Coarse Beam Alignment**

- Slide the beam block along the path of the laser, increasing the distance between the laser and the beam block. Keep the laser dot on the beam block as you increase the distance until the beam block is about 1-1.5m away.

- Slowly adjust the position and orientation of the laser while keeping the laser output on the beam block until the beam path roughly follows a table bolt-hole line.

**Fine Beam Alignment**

- Place a wooden ruler square on the table so that the short side is flush against the table and the long side is perpendicular to the plane of the table.
    - **SAFETY NOTE:** do not use a plastic or metal ruler to do this. Plastic and most metal rulers have a smooth reflective surface. The uncontrolled laser reflection from the surface of a plastic or metal ruler is a safety hazard. Wood rulers do not have a reflective surface.

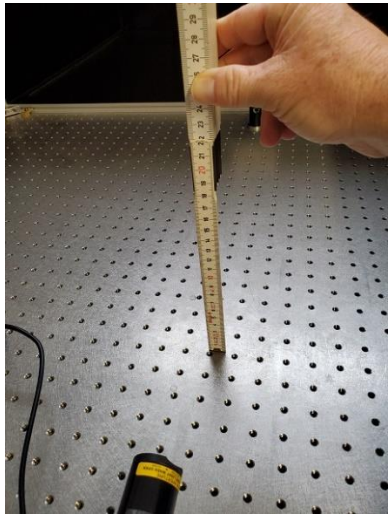

- **"Near" ruler placement.** Pick a table bolt hole along the desired bolt hole beam path near (~15cm) the output of the laser. Move the ruler so that the corner of the ruler is centered exactly on the far edge of the table bolt hole.

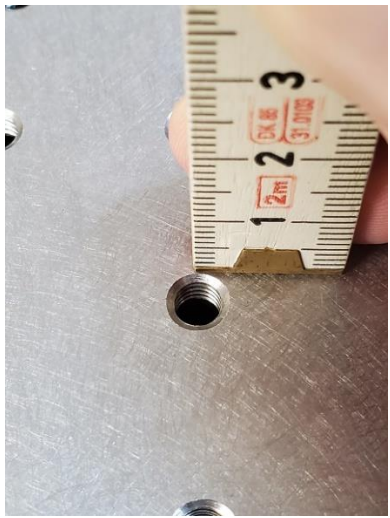

- **Horizontal laser movement.** Slide the laser on the table in the direction perpendicular to the beam path until exactly half of the laser dot appears on the edge of the ruler.

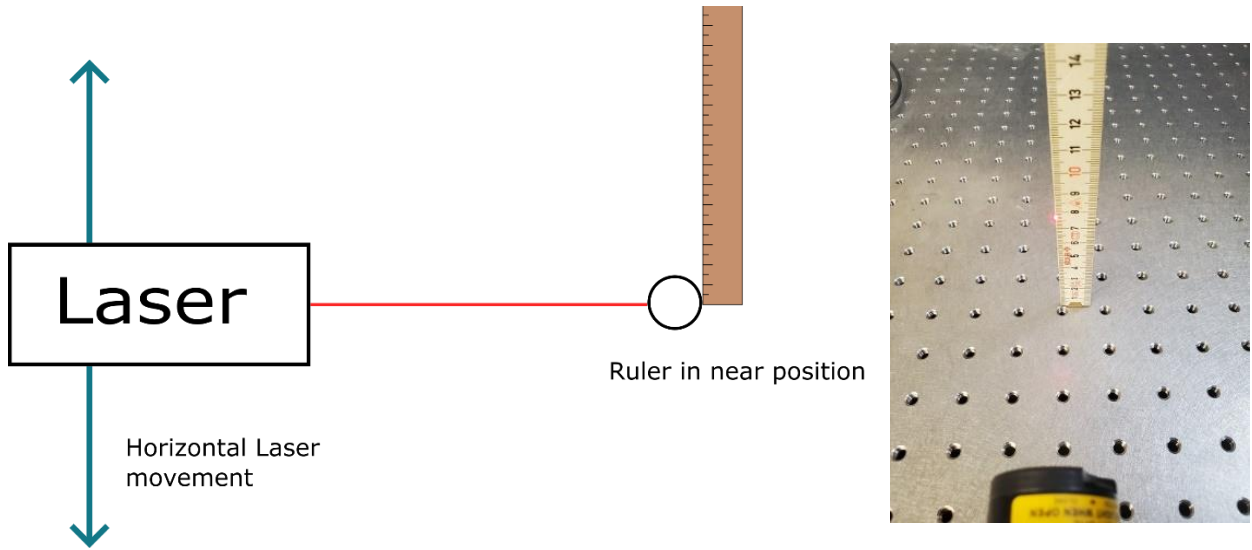

- **"Far" ruler placement.** Pick a table bolt hole along the desired bolt hole beam path far from the output of the laser. "Far" here is a relative term.
    - If you are doing this by yourself, pick a distance where you can comfortably hold the ruler with one hand and use your other hand to carefully and accurately rotate the laser. **Remember laser safety.** You don't want your arms so far apart that your head is lowered such that your eye level approaches the level of the laser beam. For a single person this distance can be ~80-110cm.
    - If you have a partner, pick a distance a bit further away, (~110-150cm). One person will hold the ruler, and the other will adjust the laser.
- Place the corner of the ruler so that it is centered exactly on the far edge of the table bolt hole.
- **Rotational laser movement.** Rotate the laser about an axis which is normal to the surface of the table and the center of rotation is near the output end of the laser. Slowly and carefully make minor rotational adjustments until exactly half of the laser dot appears on the edge of the ruler.

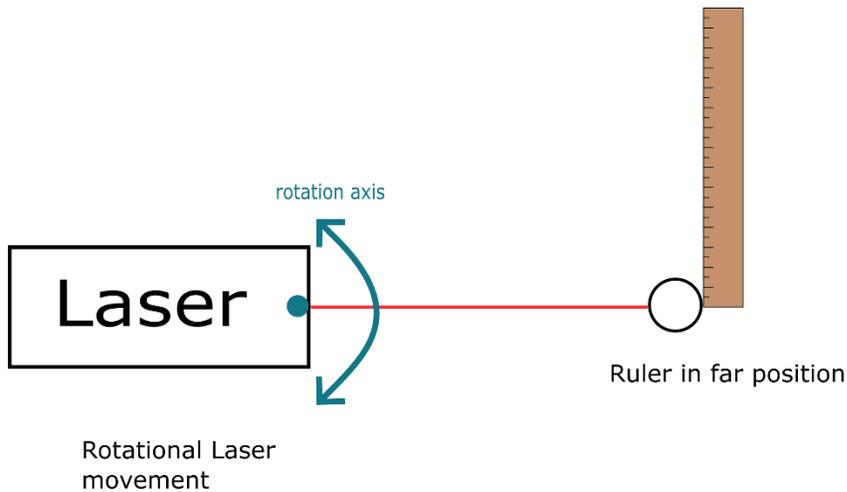
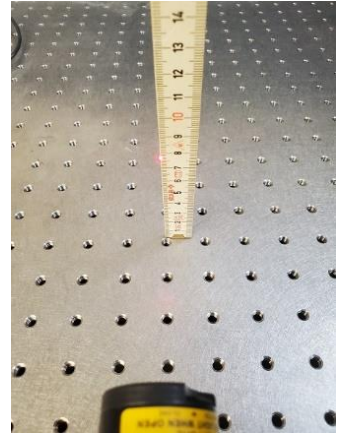

Rotational Laser movement · Ruler in far position

- **Repeat** the "near" ruler placement with horizontal adjustments and the "far" ruler placements with rotational adjustment **until the beam is perfectly cut in half by the ruler at both locations.** This is an iterative process. However, with each round of adjustments you should need less and less movement of the laser. In other words, you are slowly walking the beam path closer and closer to your desired line with each round of adjustments.

**Secure the Laser**

Once the laser is aligned perfectly to follow a table bolt hole line. Secure the laser to the table using the "How to Secure Optical Hardware to an Optics Table" tutorial at the end of this document. **Recheck the beam alignment** after securing the laser to the table. If the beam shifted during the clamp tightening process, loosen the clamps, realign the laser using the above method, and tighten the table clamps again.

**Beam Height Variation**

Ideally, the height of the beam above the table should also be constant, i.e., the millimeter location of the laser dot on the ruler at *both* the near and far locations is the same. Most laser housings and mounts do not allow for adjustment of the vertical angle of the laser. If this is the case, then you are done. If vertical adjustment of the laser output, or laser housing, is available then adjust the output of the laser such that the height of the dot on the ruler at both the near and far locations is the same.

**Final Setup**

The final setup should look like:

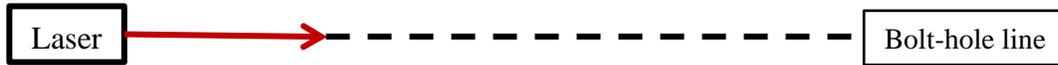

## What is next?

Now that you have completed the task of aligning the output of the laser beam to follow a chosen bolt-hole line in the table, the following tutorials start with this setup:

- Tutorial 2: horizontal shift along parallel table bolt-hole lines
- Tutorial 5: vertical shift without polarization rotation
- Tutorial 6: vertical shift with 90* polarization rotation
- Tutorial 8: align a lens to a beam path

# Tutorial 2: Horizontal shift along parallel table bolt-hole lines

**Initial Setup**

The initial setup for this tutorial should be the final setup of **Tutorial 1: laser output along a table bolt-hole line.** The output of the laser should exactly follow one of the bolt-hole lines of the optics table. If not, adjust the position and angle of the laser (according to the previous tutorial) such that the laser path exactly follows a chosen line of the optics table.

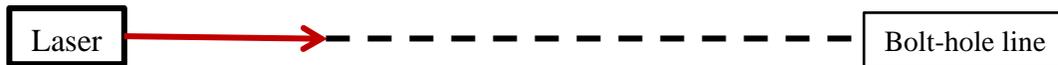

**Prerequisites**

Before beginning this tutorial, the student must have either:

- completed Tutorial 1: laser output along a table bolt-hole line, or
- measure and verify that the output of the laser is propagating *exactly* above a table bolt-hole line

**Tutorial Goals**

The goals of this tutorial are to:

- learn how to shift a beam horizontally such that the final beam path is parallel to the original beam path and *exactly* follows above a different bolt-hole line in the optics table
- insert two mirrors into the beam path so that the laser can be precisely walked to a desired new beam path

**Equipment**

- laser
- ability to clamp the laser to the table
    - some laser housings allow for them to be set on the table with edges available for table clamps, or
    - laser housing with an M6 thread on the bottom
        - M6 threaded rod
        - post
        - post holder

- base
- table clamps and bolts
- Wooden ruler, ~30 cm
- Beam block
- Two mirrors on kinematic mounts, associated posts, post holders, bases, and table clamps

**Final Setup**

The final setup you are working toward should eventually look like:

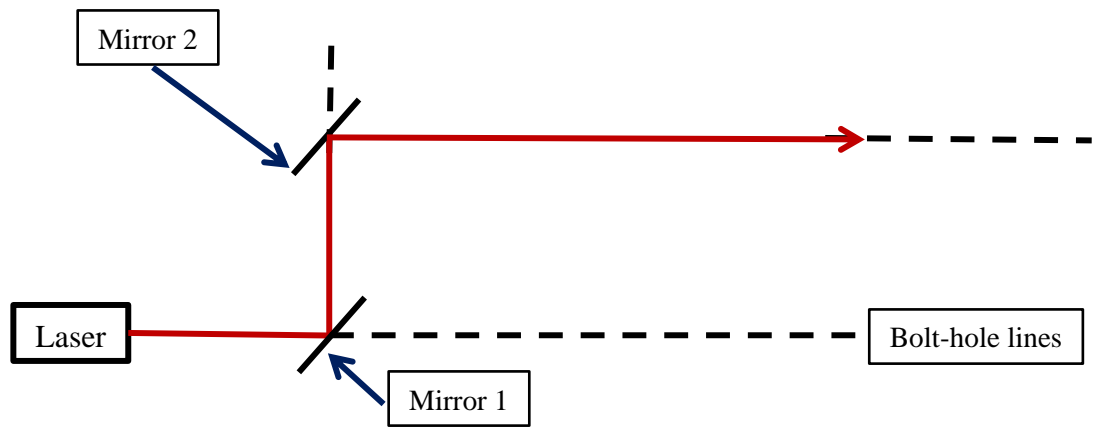

Carefully follow the outlined procedures to get there.

**Procedure**

1. Place a beam block near the output of the laser (approximately 20-50 cm from the output)
2. Turn on the laser and verify that the beam terminates on the beam block
3. Use a *wooden ruler* to measure the height of the laser dot on the beam block above the table, and record this value.
    - **SAFETY NOTE:** do not use a plastic or metal ruler to do this. Plastic and most metal rulers have a smooth reflective surface. The uncontrolled laser reflection from the surface of a plastic or metal ruler is a safety hazard. Wood rulers do not have a reflective surface.
    - **Additional reflective surfaces caution:** avoid placing any other smooth reflective surfaces in or near your working area, on or above the table, with the laser beam; i.e., screw drivers, ball drivers, watches, rings with flat surfaces, etc. If you wear a watch, it is good optics lab protocol to take your watch off if your hands are going to be anywhere near the laser beam.
4. Turn off the laser

5. Find, or assemble, two mirrors mounted in a kinematic mount.  Kinetic mounts allow for the precise movement, or adjustment, of an optical component's horizontal and vertical angle.  They are easily identified by the two adjustment knobs on the back.

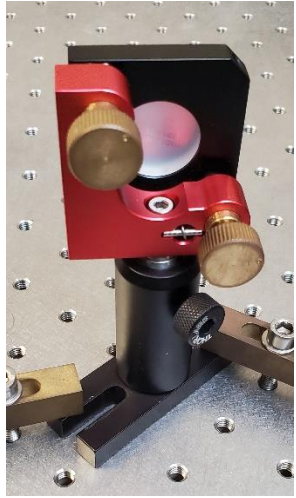

- **HANDLING OPTICS NOTE:** Never touch the surface of an optic with your bare fingers!  Always either:
    - **Use gloves (preferably)**
    - or only very carefully handle the objects by their non-coated edges
    - In the case of mirrors you can also handle/touch the non-coated back surface.
6. Verify, or adjust, the kinematic mount such that the adjustment screws are in the center of their travel.  When the screws are in the center of their travel, the fixed and movable plates should be parallel to each other.

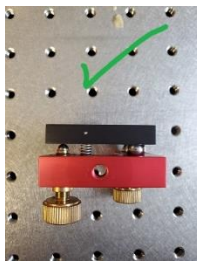 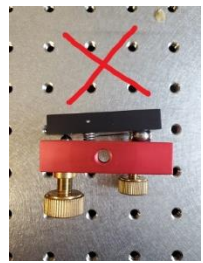

7. Attach the kinematic mount to an appropriate post, post holder, and base so that the distance from the bottom of the base to the center of the mirror will be about the beam height you previously measured.
    - **Good optical technique tip:** when measuring the height or position of the mirror, *the ruler should never touch the surface of the mirror*.  Place the ruler such that

the edge of the ruler is measuring the height of the outermost horizontal edge of the mirror.

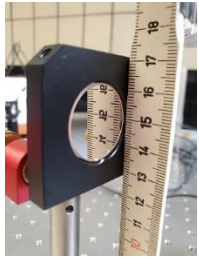

8. **Align the surface of the mirror to be at 45° to the table lines directly above a bolt hole**
    - Pick a bolt hole where you want the beam to follow a perpendicular, or orthogonal, line of bolt holes in the table.
    - Place the mirror assemble on the table such that when you look down at the hole from above the surface of the mirror is along your line of sight and the mirror covers half of the bolt hole below.  In other words, you want your line of sight to be from the center of the hole, normal to the table surface, along the vertical center of the surface of the mirror, to your eye.  This should get the surface of the mirror close to being directly above the chosen bolt hole over which you plan to change the direction of the laser beam.  See Figs below.

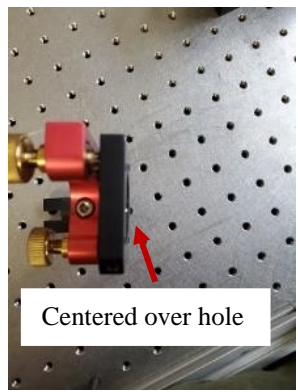

Centered over hole

    - Now rotate the mirror assembly about an axis which is the previous line of sight along the surface of the mirror until the mirror surface is at 45° to the table lines *and* the mirror surface is directly above the chosen bolt hole.

**How to visually check that a mirror surface is exactly 45° to the table lines *and* centered above a chosen bolt hole.**

- Close one eye and move your head such that you are horizontally looking along the same bolt-hole line as the original path of the beam before the mirror, and your line of sight is about 20-30° above the table. The table line of the original beam path should be the only line that appears to be going straight away from you. The adjacent bolt-hole lines should appear to be converging towards your chosen line.
- Now look at the bolt-hole line in the reflection of the mirror. You should see two sets of lines now, one set which is the table lines themselves, and one set in the reflection of the mirror.
- The mirror surface is at 45° to the table line pattern when the pattern in the reflection exactly lines up with the pattern observed directly from the table.

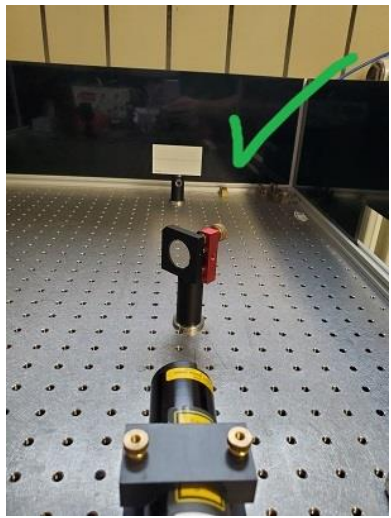 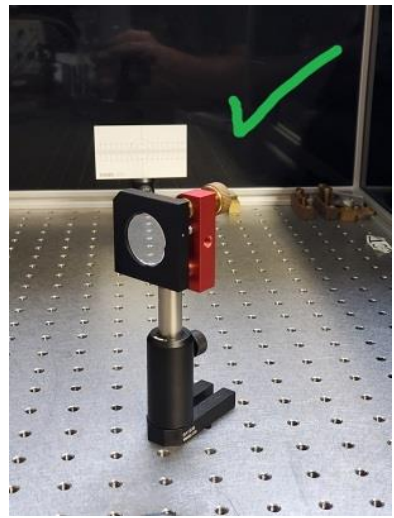

Image of bolt-hole line in the mirror aligns with table bolt-hole line

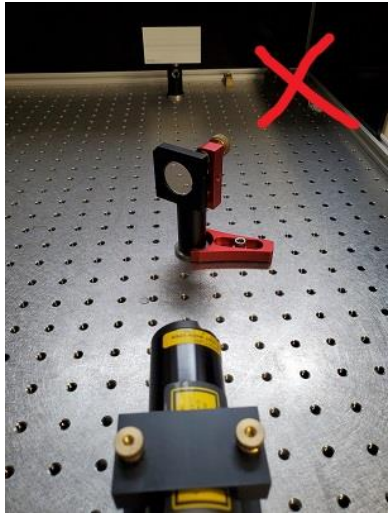

Image of bolt-hole line in the mirror does not align with table bolt-hole pattern below.

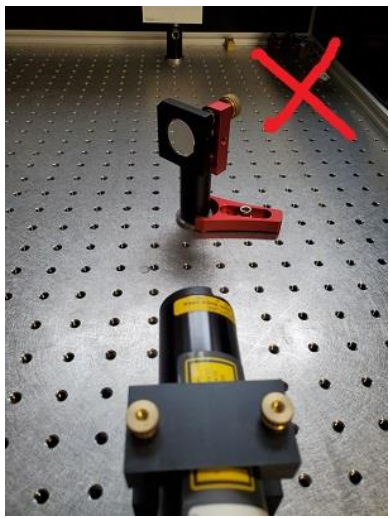

Image of line in the mirror is parallel but not co-linear with bolt-hole line of the beam path. This occurs when the mirror surface is at 45°, but the surface is not centered above the center of the bolt hole.

9. Secure the mirror to the table in this position.
    - See "How to Secure Optical Hardware to an Optics Table" at the end of Tutorial 1.
10. Move the beam block to the chosen bolt-hole line after the first mirror
11. Turn on the laser and verify that the laser beam is now following a path perpendicular to the original path and along a table bolt-hole line.
12. Turn the laser off (or close the laser shutter, or move the beam block to the output of the laser)

13. Choose a bolt hole where you want to place the second mirror such that the final beam path is parallel and horizontally offset from the original beam path.
14. Follow the same procedures as the first mirror to place, and secure, the second mirror such that the mirror surface is centered above the chosen bolt hole and is exactly 45° to the table lines.
15. Move the beam block so that it is some distance (say 30-50 cm) past the second mirror.
16. Turn on the laser
17. Verify that the beam is incident upon the approximate center of each mirror and terminates on the beam block

**Final beam path adjustment**

Here we want to fine tune the final path of the beam so that it exactly follows a bolt-hole line using a "near-near far-far" method.

**Note:** if the original path of the laser exactly follows a bolt-hole line, and the two mirrors are each centered above a bolt hole with their surfaces at 45° to the table bolt pattern then the final path of the beam should be very close to the line desired. If the final beam path does not *exactly* follow the desired bolt-hole line use the near-near far-far method to adjust the beam path so that it does.

**Near-near far-far method**

**Note:** "near" refers to a relative location on the beam path *near* the output of the laser. "Far" refers to a relative location on the beam path *far* from the laser.

- Pick a "near point" and a "far point" along the final path of the laser beam. Each point should be directly above a bolt hole in the table with the near point about 10-12.5 cm from the second mirror, and the far point about 30-40 cm past the near point.

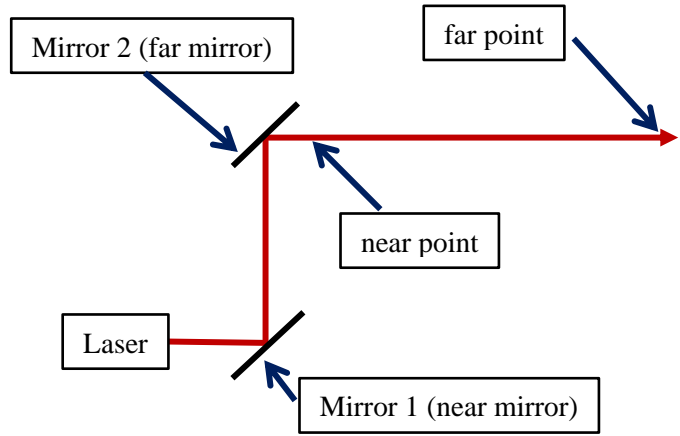

- **Near-near**:
    - Place the wooden ruler so that its edge is centered on the edge of the near point bolt hole.  (This is the same method as used in Tutorial 1).

    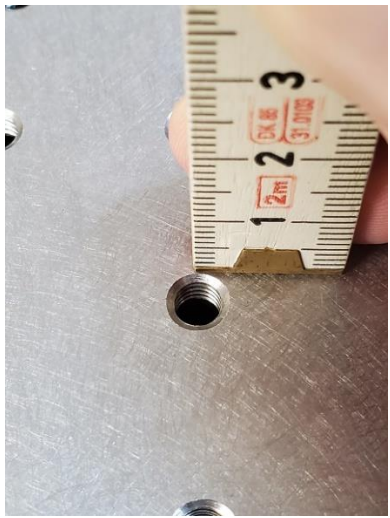

    - Use the two adjustment knobs on Mirror 1 ("near" mirror) to steer the beam such that it is cut in half by the edge of the ruler.
    - Note the height of the beam on the ruler.
- **Far-far:**
    - Place the wooden ruler so that its edge is centered on the edge of the far point bolt hole.
    - Use the two adjustment knobs on Mirror 2 ("far" mirror) to steer the beam such that it is cut in half by the edge of the ruler.
    - Note the height of the beam on the ruler.  If it is not exactly the same height as it was at the near point, adjust the vertical knob on Mirror 2 until the height of the beam is the same.

- Go back to the near point and check the beam. (By adjusting Mirror 2, the beam's position at the near point will have changed a bit. The near-near far-far method is an iterative process that may take some time. However, you should find that you are slowly converging to a desired path for both positions.)
- Repeat the near-near and far-far alignment process until the beam *exactly* follows the desired bolt-hole line and is parallel to the table.

## What is next?

Now that you have completed the task of an optical setup capable of steering the output of a laser onto a desired path, the following tutorials start with this setup:

- Tutorial 3: walk a beam horizontally and vertically
- Tutorial 4: align a second laser beam along an established beam path
- Tutorial 7: align the beam through a cage system
- Tutorial 8: align a lens to a beam path

# Tutorial 3: walk a beam horizontally and vertically

**Initial Setup**

The initial setup for this skill should be the final setup for Tutorial 2: horizontal shift along parallel table bolt-hole lines.

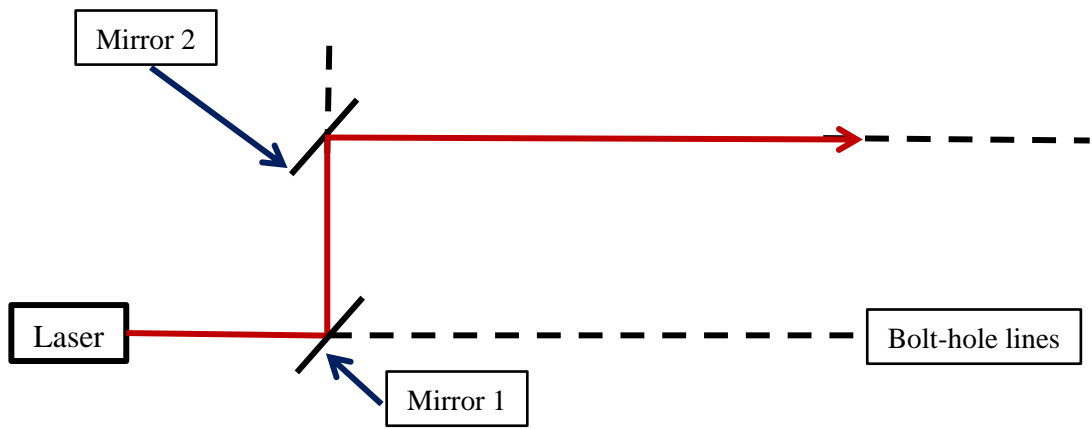

**Prerequisite**

The prerequisite for this tutorial is the successful completion of Tutorial 2.

**Tutorial Goals**

The goals of this tutorial are to:

- learn how to center an iris on a beam
- learn how to establish a given beam path using two irises
- learn how to walk a laser beam, both horizontally and vertically, using the near-near far-far method

**Equipment**

- Equipment necessary for Tutorial 2
- Two irises and associated posts, post holders, bases, and table clamps

**Final Setup**

The final setup you are working toward should eventually look like:

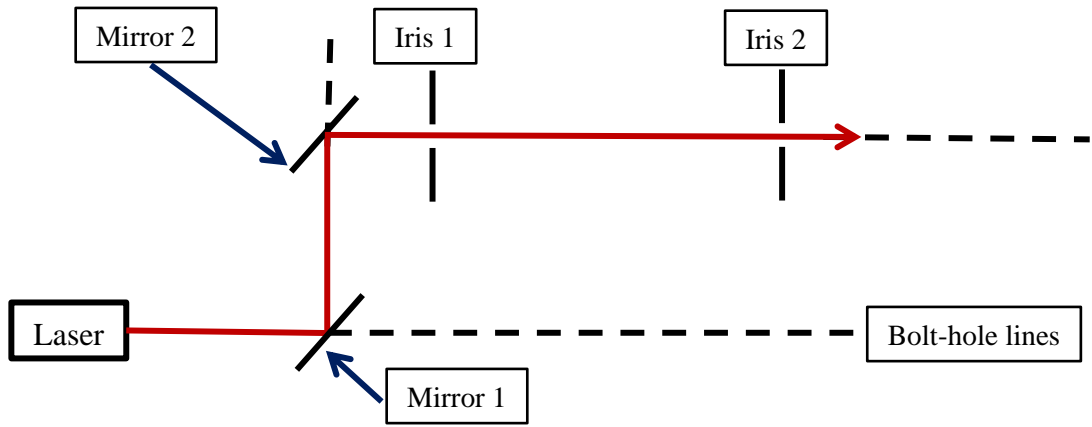

Carefully follow the outlined procedures to get there.

**Procedure**

**Part 1: Establishing the initial beam path**

18. If the initial setup is not assembled, or complete then complete the procedures for Tutorial 2.
19. Assemble an iris on a post, post holder and base.
20. Place the iris on the beam path near (about 12.5 to 15 cm) Mirror 2 and roughly centered on the beam. In other words, when the iris is open roughly halfway the beam passes through the iris. Visually check that the iris is centered on the beam.
21. Place a viewing screen after the iris so that the entire beam profile is observed.

**Visually checking if an iris is centered on a beam using the beam shadow on the viewing screen**

- You may need to dim the lights to see the lower intensity regions away from the center of the beam.
- Open and close the iris and observe the shadow of the beam relative to the beam profile.

- An iris is centered if the shadow of the iris is *perfectly* concentric with the beam profile.  As the iris is closed the center of the shadow should also be the bright central spot of the beam.

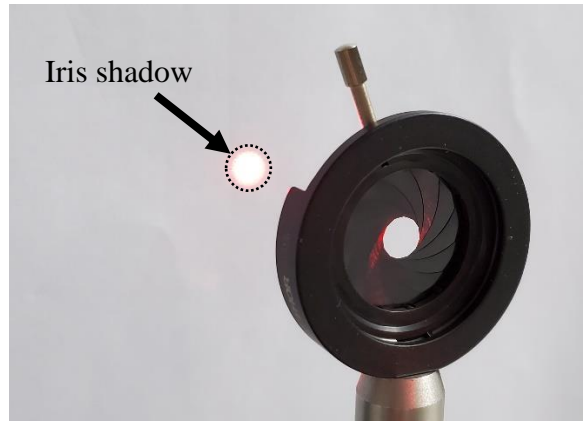

Iris shadow

**Visually checking if an iris is centered on a beam using the beam incident on the iris diaphragm**
- Close the iris as much as possible.
- **If the size of the beam is much smaller than the minimum iris setting** (not ideal), you must also use the edge of a piece of paper or a business card.
    - place the paper edge so that is *horizontal* across the opening of the iris and the edge is exactly across the widest part of the iris.

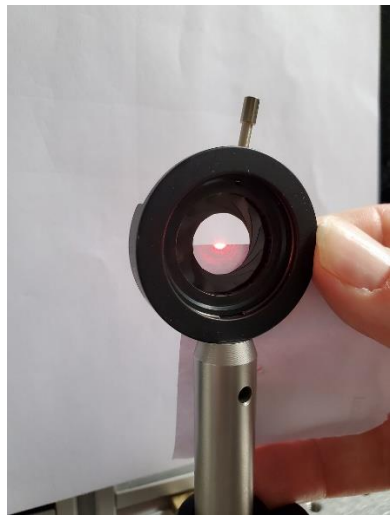

   - Carefully adjust the *height* of the iris assembly using very small movements until the beam is cut in half by the horizontally held paper.  Retighten the post set screw to hold the post at the correct height.
   - place the paper edge so that it is *vertical* across the opening of the iris and the edge is exactly across the widest part of the iris.

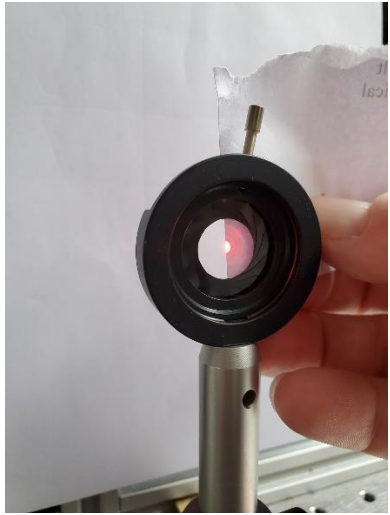

- Carefully adjust the *horizontal* placement of the iris assembly (by loosening the bolts holding it to the table and sliding the assembly ever so slightly on the table horizontally). Retighten the bolts holding the assembly to the table.
- **If the size of the beam is larger than the minimum iris setting** (ideally sized iris):
    - Close the iris down until you can just barely start to see a reflection on the inner rim of the iris.
    - Losen the bolts holding the iris assembly to the table and carefully adjust the horizontal position of the iris until the reflections of the beam on the horizontal edges of the inner iris rim, at the widest part of the iris, are symmetric. Retighten the bolts holding the iris assembly to the table.

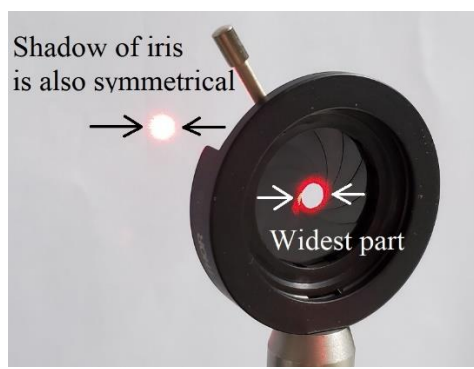

    - Losen the set screw on the post of the iris and carefully adjust the vertical position of the iris until the reflections of the beam on the vertical edges of the iris inner rim are symmetric. Retighten the post set screw.

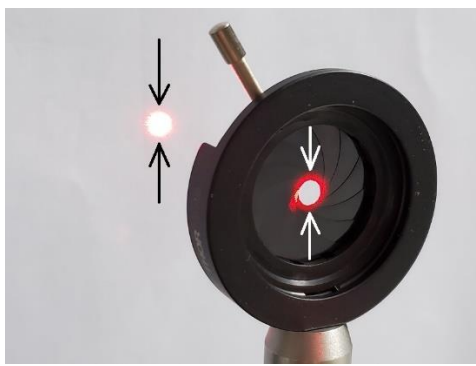

22. Using small and careful adjustments and alternate between observation of the beam shadow on the viewing screen and observation of the beam incident upon a fully closed iris, move the iris assembly horizontally until the iris is horizontally centered on the beam.
23. Secure the iris assembly to the optics table.  See Tutorial 1.
24. Loosen the set screw on the post holder and adjust the vertical position of the iris until it is perfectly centered on the beam.  Use small and careful adjustments and alternative between observation of the shadow of the beam on the viewing screen and observation of the beam incident upon a fully closed iris.
25. Tighten the set screw on the post holder to secure the vertical placement of the iris assembly.
26. After tightening the table clamps, and tightening the post set screw, check that the iris is still centered and did not shift during the securing and tightening process.
27. Assemble another iris, post, post holder, and base.
28. Place this iris in the beam path far from Mirror 2 (about 30-50 cm past Mirror 2).
29. Using the same centering and visual checking procedures as before, center the far iris on the beam path.

Notify your instructor that you have completed Part 1.

**Part 2: Walking a beam horizontally and vertically**

Your instructor just moved the irises both horizontally and vertically.

**General note:**  When walking a beam both horizontally and vertically only one direction should be adjusted at a time.  Additionally, only one mirror should be adjusted at a time.

1. Use the near-near far-far method (See Tutorial 2) to walk the beam to be horizontally centered on each iris.
    a. Close the near iris.
    b. Adjust the horizontal knob on the near mirror (Mirror 1) to move the beam observed on the diaphragm of the near iris until the beam is horizontally centered.
    c. Open the near iris and close the far iris.
    d. Adjust the horizontal knob on the far mirror (Mirror 2) to horizontally center the beam observed on the diaphragm of the far iris.
    e. Close the near iris and recenter the beam horizontally using the near mirror.
    f. Repeat near and far horizontal adjustments until the beam is horizontally centered on both irises.
2. Use the near-near far-far method to walk the beam vertically until the beam is vertically centered on each iris.
3. Double-check that the beam is now centered on each iris using a viewing screen after each iris as in Part 1.

## What is next?

Now that you have completed the task of learning the near-near far-far method to walk a laser onto a desired path, the following tutorials start with this setup:

- Tutorial 4: align a second laser beam along an established beam path
- Tutorial 7: align the beam through a cage system
- Tutorial 8: align a lens to a beam path

# Tutorial 4: align a second laser beam along an established beam path

**Initial Setup**

The initial setup for this skill should be the final setup for **Tutorial 3: Walk a beam horizontally and vertically**.

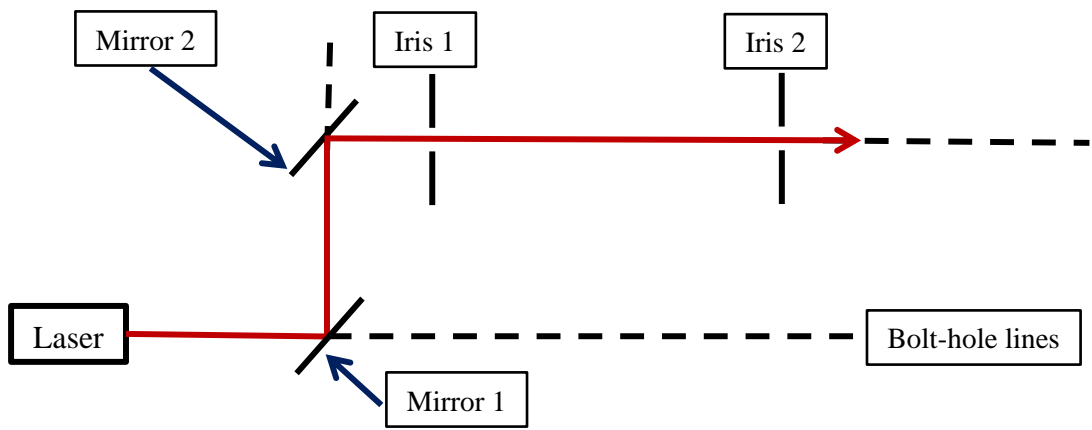

**Prerequisite**

The prerequisite for this tutorial is the successful completion of Tutorial 3.

**Tutorial Goals**

The goal of this tutorial is to have two different laser sources where the user can quickly and easily switch between the two laser sources.  However, the challenge arises from the desire for the final beam path of each laser source to be identical.  The flipper mirror allows for the output to be a single identical path for either laser beam. In other words, the final output of this setup is a single laser beam path with a selectable laser source.

**Equipment**

- Equipment necessary for Tutorial 2 or 3 shown above
- Second laser
- Mirror, mount, post, post holder, table clamp
- Flipper mirror and mounting hardware

**Final Setup**

The final setup you are working toward should eventually look like:

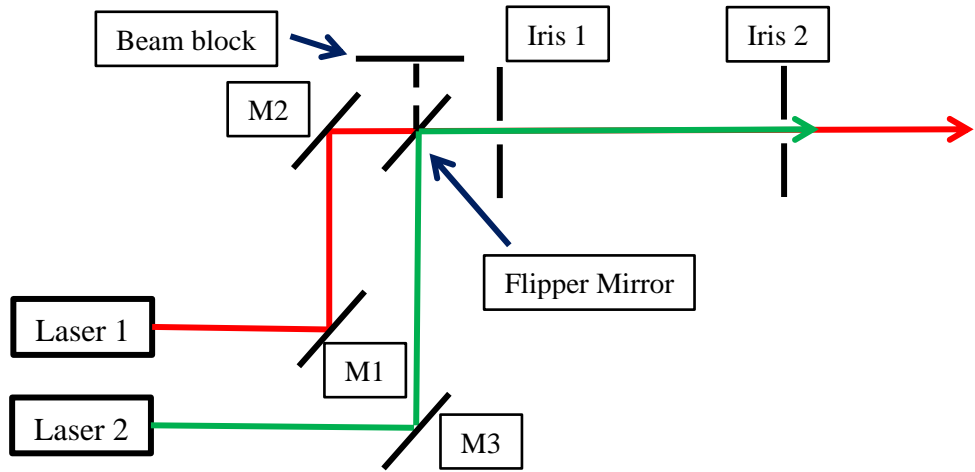

Carefully follow the outlined procedures to get there.

**Procedure**

Overall, the procedure for this tutorial is similar to combining Tutorials 1, 2, and the second part of 3 for the second laser source to match the beam path of the first laser source.

1. Setup the second laser, Laser 2, such that its initial beam path follows a bolt-hole line parallel to the first laser, Laser 1. Choose a parallel path which is about 7.5 to 15 cm from Laser 1. (See Tutorial 1)

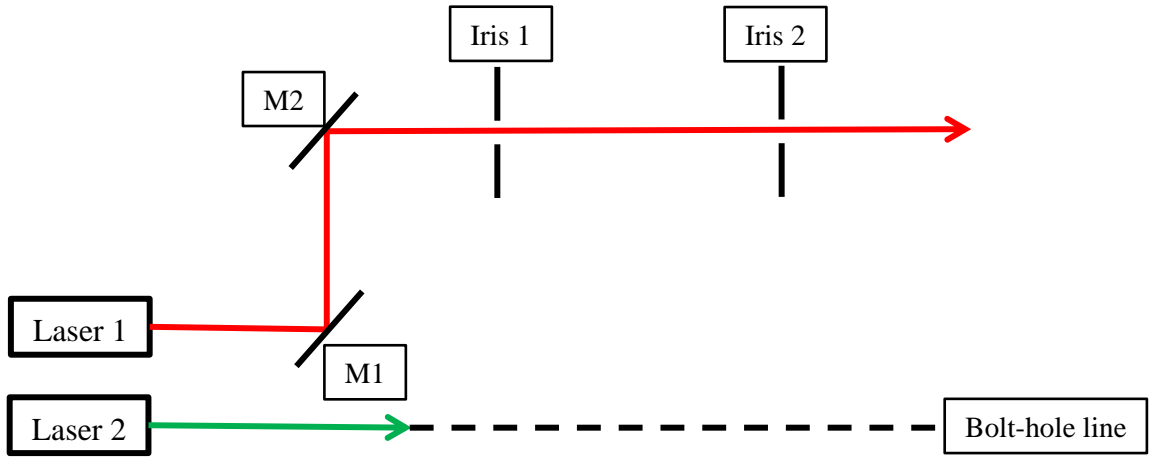

2. Choose another bolt-hole line parallel to the line between M1 and M2 and about 7.5 to 10 cm from it.
3. Assemble a mirror and place it in the path of Laser 2 such that it reflects the beam *exactly* at 90° above the bolt hole connecting the two perpendicular bolt-hole lines. Double-check that the beam path of Laser 2 is directly above a bolt-hole line and is 7.5 to 10 cm from the path between M1 and M2. (For placing the mirror at *exactly* 45° to the table lines *and* centered above a specific bolt hole see Tutorial 2).
    a. **HANDLING OPTICS NOTE:** Never touch the surface of an optic with your bare fingers! Always either:
        i. **Use gloves (preferably)**
        ii. or only very carefully handle the objects by their non-coated edges
        iii. In the case of mirrors you can also handle/touch the non-coated back surface.

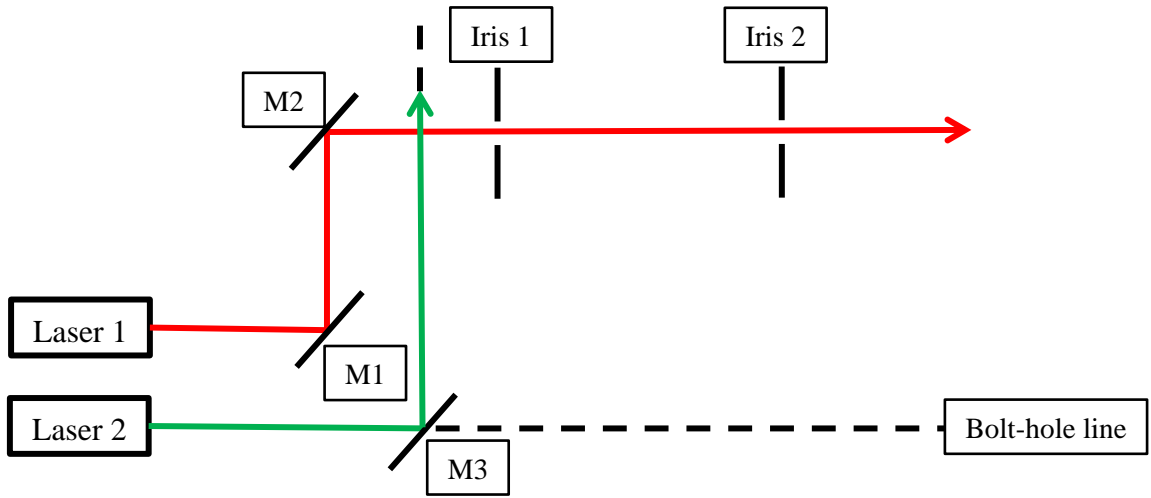

4. Place a beam block some distance past the path between Mirror 2 and the irises to safely terminate Laser 2.
5. Check the location of Iris 1 and adjust if necessary. Iris 1 should be about 7.5 to 10 cm beyond the beam line after Mirror 3. If Iris 1 is closer than this then move the iris so that it is about 7.5 to 10 cm away and carefully recenter the iris on the beam from Laser 1 and secure it to the table with a table clamp.
6. If you did move Iris 1, and if space allows, move and recenter Iris 2 and equal amount such that the separation distance between the two is the same as before moving Iris 1.
7. Assemble the flipper mirror into a kinematic mount on a post in a post holder and base.
8. Using the same procedure as with Mirror 3, place the flipper mirror into the beam path such that its surface is 45° to the table bolt hole lines, and is perfectly centered above a bolt hole.
9. Flip up the flipper mirror and use the near-near far-far method to walk the beam of Laser 2 such that it is perfectly centered on each iris. In this case the "near" mirror is Mirror 3 and the "far" mirror is the flipper mirror. (See Tutorial 3 for walking a beam).
10. Flip the flipper mirror down and double-check that the irises are perfectly centered on the beam from Laser 1.
11. Carefully and slowly flip the flipper mirror up and double-check that the path of Laser 2 is centered through each iris.

## What is next?

Now that you have completed the task of aligning two laser beams onto a single path the following tutorials start with this setup and can lead to experiments where you can choose the laser source with a simple flip of the flipper mirror:

- Tutorial 7: align the beam through a cage system
- Tutorial 8: align a lens to a beam path

# Tutorial 5: vertical shift without polarization rotation

**Initial Setup**

The initial setup for this skill should be the final setup for **Tutorial 1: laser output along a table bolt-hole line**.

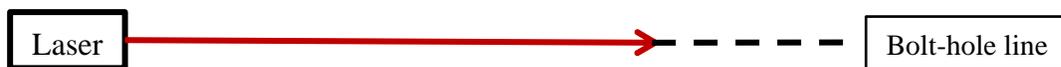

**Prerequisites**

The prerequisite skills for this tutorial are to be able to mount mirror into the beam path such that the reflected beam is *exactly* perpendicular to the incident beam, and to be able to steer the resultant beam both horizontally and vertically onto a desired final beam path. Tutorials for these skills are in:

- Tutorial 2: horizontal shift along parallel table bolt-hole lines
- Tutorial 3: walk a beam horizontally and vertically

**Tutorial Goals**

The goal of this tutorial is to shift a laser beam path vertically using two mirrors and 90° reflections where the output beam follows the same bolt-hole line as the initial beam, but is vertically shifted using a pair of mirrors.

**Equipment**

- Setup Tutorial 1 (Laser with table clamps, beam block)
- Wooden ruler, ~30 cm
- Two mirrors on kinematic mounts, right-angle clamps, associated posts, post holders, bases, and table clamps

**Final Setup**

The final setup you are working toward should eventually look like:

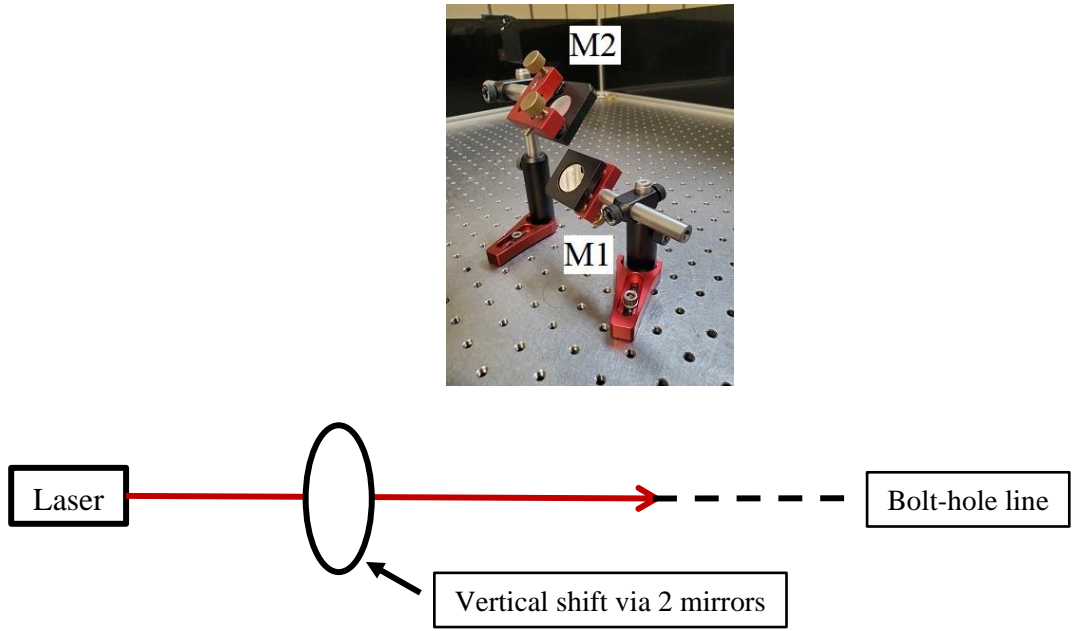

Carefully follow the outlined procedures to get there.

**Procedure**

1. Make sure that you have a beam block near the output of the laser
2. Turn on the laser and measure the height of the laser beam.
3. Turn off the laser or close the shutter.

**Assembling Mirror 1 to reflect the beam from parallel to the tabletop to normal to the table**

4. Assemble a mirror on a kinematic mount.

    - **HANDLING OPTICS NOTE:** Never touch the surface of an optic with your bare fingers!  Always either:
        - **Use gloves (preferably)**
        - or only very carefully handle the objects by their non-coated edges

- In the case of mirrors, you can also handle/touch the non-coated back surface.

5. Using two posts, a right angle clamp, a post holder, base and table clamp assemble the following:

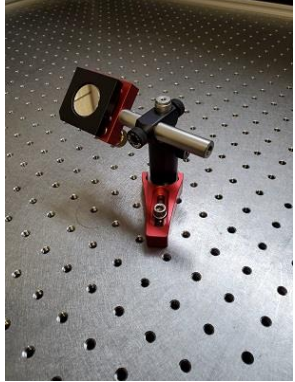

6. Make sure that the two adjustment screws on the kinematic mount are set such that they are in the middle of their travel and that the mirror mount plate is parallel to the mounting plate.

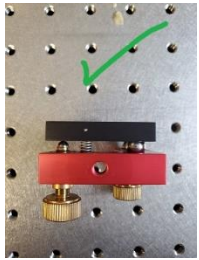 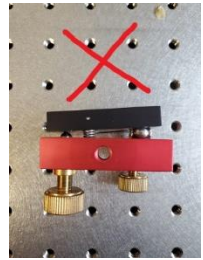

7. **Set the mirror to 45°**
- Measure the height of the center of the mirror by measuring from the table surface to the center of the mirror along the edge of the kinematic mount

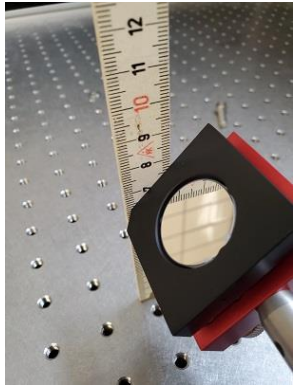

- Look *directly down* at the mirror with your eyes directly above the bolt-hole where the beam is changing direction. This should also be the center of the mirror. Use the bolt holes as your guide.

**How do you know if you are looking *directly down* at the chosen bolt hole?**

You cannot see the hole directly under the mirror, but you can see the two bolt-hole lines that intersect at the chosen hole, and sets of holes symmetrically around these lines.

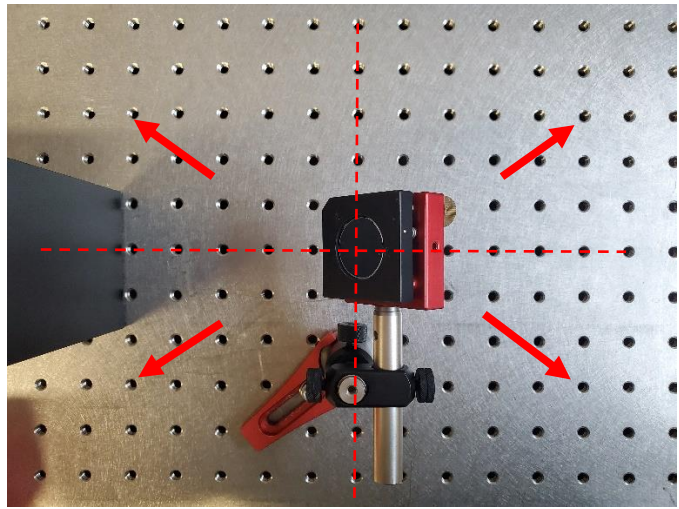

Pick a few holes that are a symmetrical distance away from directly under the mirror. Move your head around such that you are looking straight down at the mirror *and* the angles of the thread patterns within the symmetrically located holes are also symmetrical.

- While still holding your head *directly above* the mirror place the ruler some distance away from the mirror (about 15-30 cm) holding it vertical and normal to the table surface, place it where you can see the ruler in the reflection of the mirror.
- Adjust the angle of the mirror by rotating the post in the right-angle clamp until you see the same height measurement in the center of the reflection.

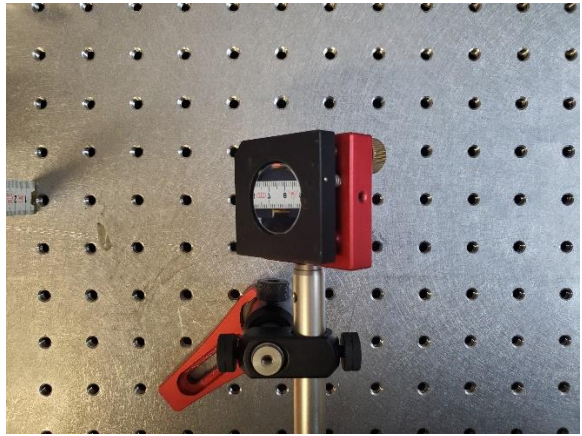

- Tighten the right-angle clamp bolt to secure the post and mirror at this angle.
- Double-check the angle visually to make sure the mirror post did not rotate during the tightening process.

**Assemble Mirror 2 to reflect a beam from normal to the table to parallel to the table**

Here we are going to assemble the second mirror of the vertical shift. This mirror will reflect a vertical laser beam back to being horizontal. The procedure is similar to assembling the previous mirror, however, there is a challenge in that you cannot look straight down at the mirror and see a reflection.

8. Assemble a mirror, kinematic mount, right-angle clamp, posts, post holder, base, and table clamp assemble as shown.

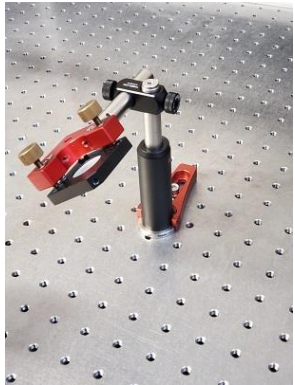

9. Loosen the thumb screw between the right-angle clamp and the vertical post in the post holder.
10. Pull the right-angle clamp off the vertical post and flip it over by 180°.

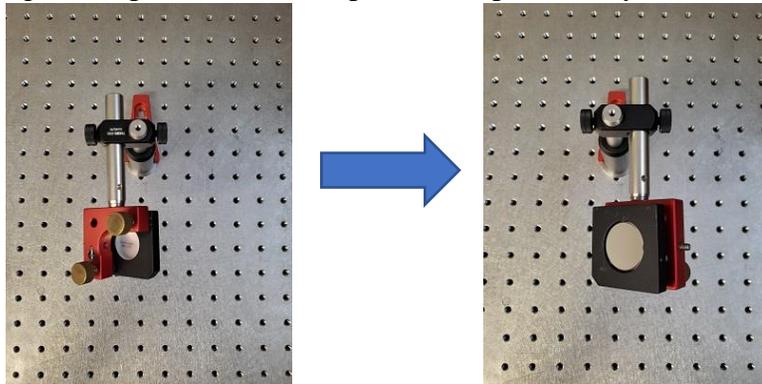

11. Now you can look straight down and see a reflection in the mirror and use the same method as before for Mirror 1. This step sets the mirror post in the right-angle clamp to exactly 45°.
12. After setting the angle and tightening the thumbscrew holding the mirror post to the right-angle clamp, loosen the thumb screw holding the right-angle clamp to the vertical post held by the post holder and flip the mirror back.

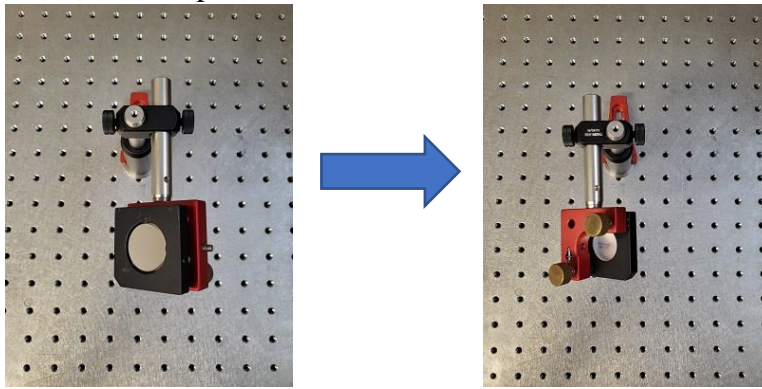

13. Place Mirror 1 in front of the output of the laser. The distance is not critical but leave some space (say 15 to 25 cm) between the laser and Mirror 1 such that other components can be placed between them; for example, power and/or polarization control.
14. Place Mirror 2 such that the center of its surface is directly above the center of the surface of Mirror 1.
15. Place a beam block right after Mirror 2.
16. Turn on the laser or open the shutter to allow the beam to enter the system.
17. Double-check that the beam safely terminates on the downstream beam block.
18. Adjust the position of Mirror 1 such that the beam is incident upon the center of the mirror. Sometimes you can see the beam itself on the mirror as a small dot. Alternatively, you can use the corner of a business card to locate the beam. **Note:** nothing should ever touch the surface of an optic except perhaps paper (i.e., the corner of a business card, etc.)

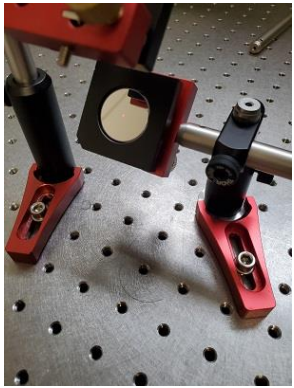 or 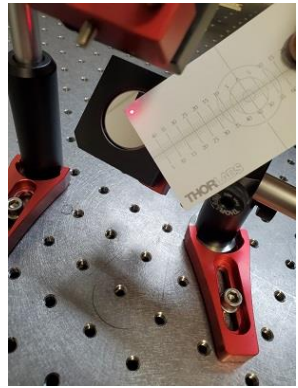

19. Slowly move the beam block farther away from Mirror 2 keeping the beam terminated on the beam block.
20. Coarsely adjust the angle of Mirror 2's assembly such that the beam after Mirror 2 roughly follows the same bolt-hole line as the beam path between the laser output and Mirror 1. Make coarse adjustments by loosening the table clamp holding the Mirror 2 assembly and carefully make very small rotation adjustments of the entire assembly.
21. Check that the beam is still incident upon the center of Mirror 2 using the corner of a business card.
22. Once the beam after Mirror 2 is roughly following the chosen bolt-hole line, tighten down the table clamp for Mirror 2.
23. Using a ruler, a near-point hole and a far-point hole, use the near-near far-far method to walk the beam such that it is parallel to the table and exactly follows the chosen bolt-hole line.

## What is next?

Now that you have completed the task of learning the near-near far-far method to walk a laser onto a desired path with a vertical shift, the following tutorials start with this setup:

- Tutorial 7: align the beam through a cage system
- Tutorial 8: align a lens to a beam path

# Tutorial 6: vertical shift with a 90° polarization rotation

**Initial Setup**

The initial setup for this skill should be the final setup for Tutorial 1.

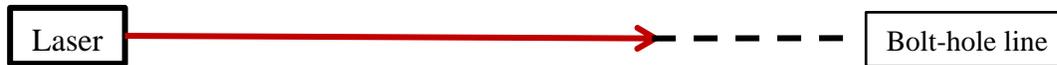

**Prerequisites**

The prerequisite skills for this tutorial are to be able to mount mirror into the beam path such that the reflected beam is *exactly* perpendicular to the incident beam, and to be able to steer the resultant beam both horizontally and vertically onto a desired final beam path. Tutorials for these skills are in:

- Tutorial 2: horizontal shift along parallel table bolt-hole lines
- Tutorial 3: walk a beam horizontally and vertically

**Tutorial Goals**

The goal of this tutorial is to shift a laser beam path vertically where the final beam path has a horizontal shift of 90° *and* the polarization has a 90° rotation. You should also be able to measure the polarization angle of the beam and verify that the rotation after the vertical shift is exactly 90°. A final setup is similar to:

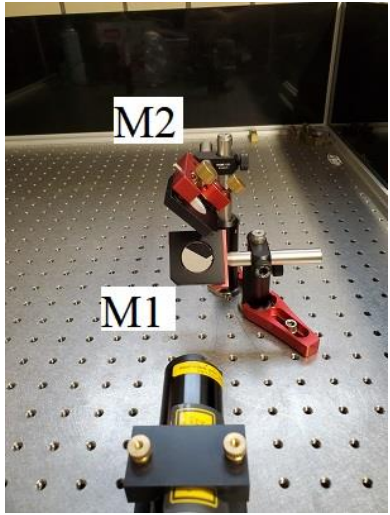

Setup for a vertical shift from a lower to a higher beam height above the table with a 90° polarization rotation. Output beam is 90° from incident and exiting to the left.

**Equipment**

- Setup Tutorial 1 (Laser with table clamps, beam block)
- Wooden ruler, ~30 cm
- Two mirrors on kinematic mounts, right-angle clamps, associated posts, post holders, bases, and table clamps

**Procedure**

Overall, the procedure for this tutorial is very similar to Tutorial 5: vertical shift without polarization rotation. The main difference is the placement of Mirror 2 which is 90° different.

24. Make sure that you have a beam block near the output of the laser
25. Turn on the laser and measure the height of the laser beam.
26. Turn off the laser or close the shutter.

**Assembling Mirror 1 to reflect the beam from parallel to the tabletop to normal to the table**

27. Assemble a mirror on a kinematic mount.

    - **HANDLING OPTICS NOTE:** Never touch the surface of an optic with your bare fingers! Always either:
        - **Use gloves (preferably)**

- or only very carefully handle the objects by their non-coated edges
- In the case of mirrors, you can also handle/touch the non-coated back surface.

28. Using two posts, a right-angle clamp, a post holder, base and table clamp assemble the following:

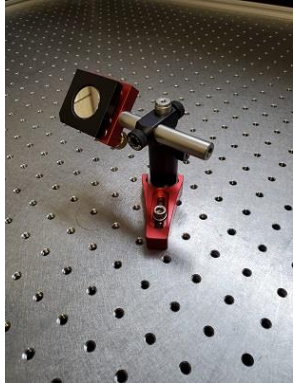

29. Make sure that the two adjustment screws on the kinematic mount are set such that they are in the middle of their travel and that the mirror mount plate is parallel to the mounting plate.

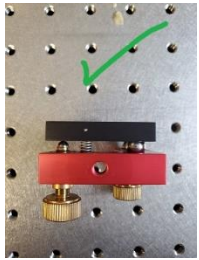 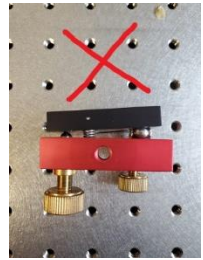

30. **Set the mirror to 45°**
- Measure the height of the center of the mirror by measuring from the table surface to the center of the mirror along the edge of the kinematic mount

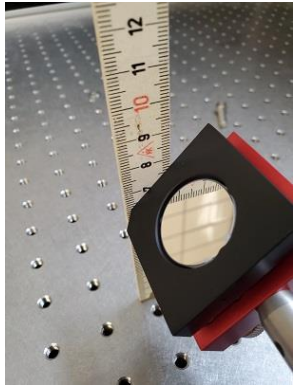

- Look *directly down* at the mirror with your eyes directly above the bolt-hole where the beam is changing direction. This should also be the center of the mirror. Use the bolt holes as your guide.

**How do you know if you are looking *directly down* at the chosen bolt hole?**

You cannot see the hole directly under the mirror, but you can see the two bolt-hole lines that intersect at the chosen hole, and sets of holes symmetrically around these lines.

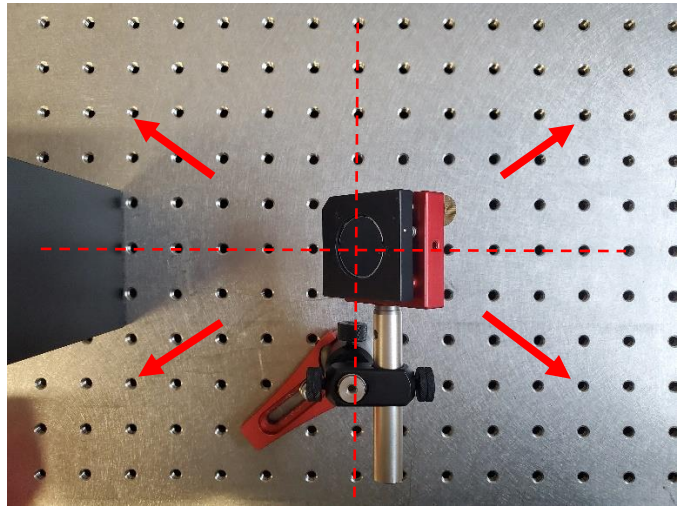

Pick a few holes that are a symmetrical distance away from directly under the mirror. Move your head around such that you are looking straight down at the mirror *and* the angles of the thread patterns within the symmetrically located holes are also symmetrical.

- While still holding your head directly above the mirror place the ruler some distance away from the mirror (about 15-30 cm) holding it vertical and normal to the table surface, place it where you can see the ruler in the reflection of the mirror.
- Adjust the angle of the mirror by rotating the post in the right-angle clamp until you see the same height measurement in the center of the reflection.

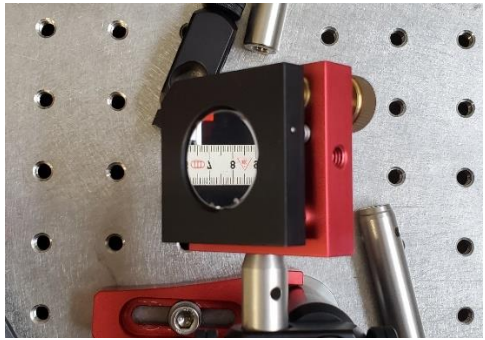

- Tighten the right-angle clamp bolt to secure the post and mirror at this angle.
- Double-check the angle visually to make sure the mirror post did not rotate during the tightening process.

**Assemble Mirror 2 to reflect a beam from normal to the table to parallel to the table**

Here we are going to assemble the second mirror of the vertical shift. This mirror will reflect a vertical laser beam back to being horizontal *and* have a horizontal shift of 90° from the incident laser beam; i.e., the final beam path will be at a 90° angle from the incident beam path. The procedure is similar to assembling the previous mirror, however, there is a challenge in that you cannot look straight down at the mirror and see a reflection.

31. Assemble a mirror, kinematic mount, right-angle clamp, posts, post holder, base, and table clamp for Mirror 2.
32. Using the orientation where incident beam path as "up", decide if you want the final beam to be deflected 90° "left" or 90° "right".

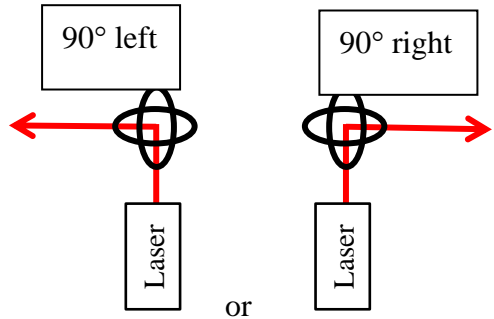

or

33. Assemble the Mirror 2 hardware as shown for either configuration:

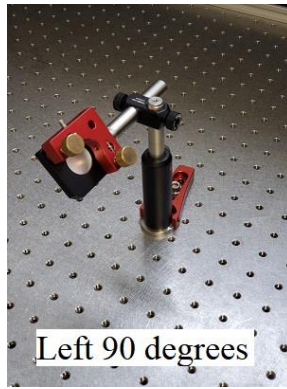
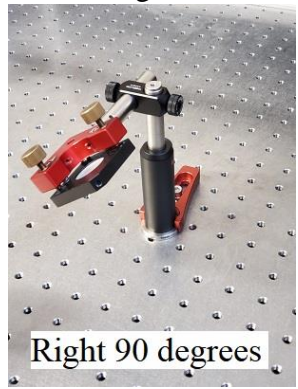

Left 90 degrees     Right 90 degrees

For the remainder of this tutorial, the "Right 90 degrees" configuration will be used as an example.

With the mirror in either Mirror 2 configuration you can not see the reflection of the mirror surface when looking straight down at it as with Mirror 1. So, in the next few steps we will flip the mirror over to be able to use the reflection to make the mirror surface 45° so that the final beam path is parallel to the table top.

34. Loosen the thumb screw between the right-angle clamp and the vertical post in the post holder.
35. Pull the right-angle clamp off the vertical post and flip it over by 180°.

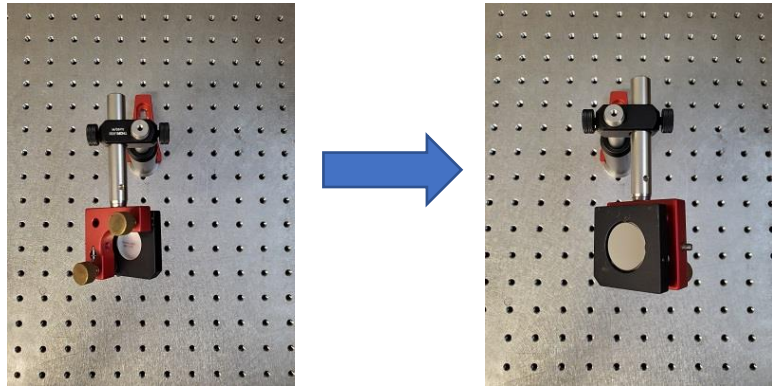

36. Now you can look straight down and see a reflection in the mirror and use the same method as before for Mirror 1.  This step sets the mirror post in the right-angle clamp to exactly 45°.
37. After setting the angle and tightening the thumbscrew holding the mirror post to the right-angle clamp, loosen the thumb screw holding the right-angle clamp to the vertical post held by the post holder and flip the mirror back.

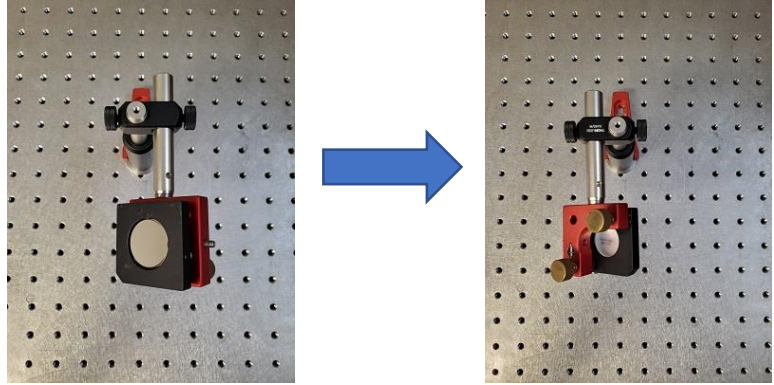

38. Place Mirror 1 in front of the output of the laser.  The distance is not critical, but leave some space (say 15 to 25 cm) between the laser and Mirror 1 such that other components can be placed between them; for example, power and/or polarization control.
39. Place Mirror 2 such that the center of its surface is directly above the center of the surface of Mirror 1 **such that the beam will be reflected at a 90° angle horizontally**. This is orientation is the key difference between this tutorial and a vertical shift *without* a polarization shift. For a left 90° vertical shift your setup would look like:

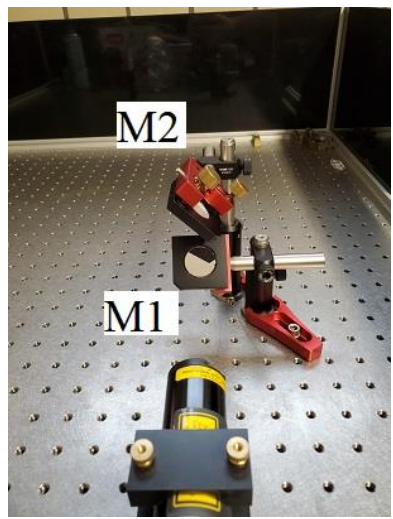

40. Place a beam block right after Mirror 2.
41. Turn on the laser or open the shutter to allow the beam to enter the system.

42. Double-check that the beam safely terminates on the downstream beam block.
43. Adjust the position of Mirror 1 such that the beam is incident upon the center of the mirror.
44. Slowly move the beam block farther away from Mirror 2 keeping the beam terminated on the beam block.
45. Coarsely adjust the angle of Mirror 2's assembly such that the beam after Mirror 2 roughly follows a bolt-hole line which is perpendicular to the beam path between the laser output and Mirror 1. Make coarse adjustments by loosening the table clamp holding the Mirror 2 assembly and carefully make very small rotation adjustments of the entire assembly.
46. Check that the beam is still incident upon the center of Mirror 2 using the corner of a business card.
47. Once the beam after Mirror 2 is roughly following the chosen bolt-hole line, tighten down the table clamp for Mirror 2.
48. Using a ruler, a near-point hole and a far-point hole, use the near-near far-far method to walk the beam such that it is parallel to the table and exactly follows the chosen bolt-hole line.

**Measure the laser polarization before Mirror 1**

49. Measure the polarization orientation of the beam between the laser and Mirror 1 using a polarizer and a power meter.
50. Place the polarizer in the beam path between the laser and Mirror 1 and place the power meter in the beam path after Mirror 2.
51. Note, and record, the position of the rotation mount's engraved scale the reading of the polarizer at maximum power transmission.

**Measure the laser polarization after Mirror 2**

52. Now place the polarizer in the beam path right after Mirror 2 and before the power meter.
53. Measure the polarization orientation of the beam after Mirror 2 and note, and record, the reading on the engraved scale of the rotation stage.
54. Verify that the two readings are 90° apart.

**Note:** if the readings are not 90° apart then the beam between Mirror 1 and Mirror 2 is not perfectly perpendicular to the beams before Mirror 1 and after Mirror 2. If this is the case, you need to walk the beam after Mirror 2 either vertically and/or horizontally until the beam between the mirrors is exactly perpendicular to the incident and final beam paths. (See "Tutorial 3: walk a beam horizontally and vertically.")

**Final setup:**

Your final setup should be similar to the following:

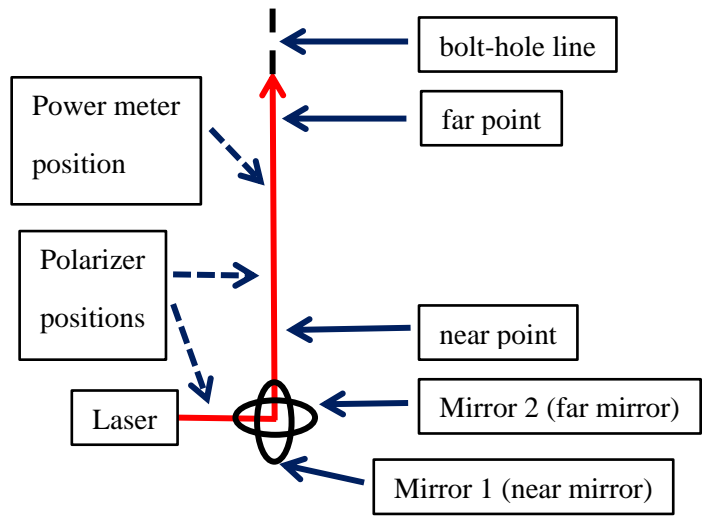

## What is next?

Now that you have completed the task of learning the near-near far-far method to walk a laser onto a desired path with a vertical shift, the following tutorials start with this setup:

- Tutorial 7: align the beam through a cage system
- Tutorial 8: align a lens to a beam path

# Tutorial 7: align a beam through a cage system

**Initial Setup**

The initial setup for this skill should be the final setup for Tutorial 2, and/or Tutorial 5 or 6.

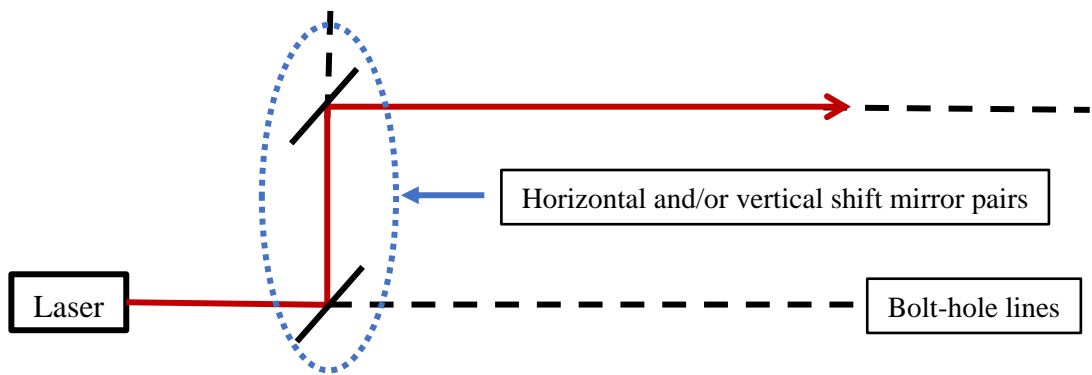

**Prerequisites**

Prior to starting this tutorial, the user should at least be proficient with horizontal shifts and walking a beam, or:

- Tutorial 2: horizontal shift along parallel table bolt-hole lines, and
- Tutorial 3: walking a beam horizontally and vertically

If the height of the laser and the height of the cage system are both fixed and not the same, then the user should also be proficient with vertical shifts, or:

- Tutorial 5: vertical shift without polarization rotation, and/or
- Tutorial 6: vertical shift with 90° polarization rotation

**Tutorial Goals**

The goals of this tutorial are to learn how to:

- assemble a cage system
- secure a cage system to the optics table using a few different methods
- coarsely align the laser beam through the cage system
- walk the beam horizontally and vertically so that is it *perfectly* centered through the cage system.

**Assumptions for this tutorial**

This tutorial will assume that the cage system is initially not built and that the user is free to set the location and height of the cage system.

If the cage system is already part of an optical setup and its height, or position, cannot be changed then the user would need to horizontally and/or vertically shift the incoming laser beam path to match the center of the cage system using skills from Tutorials 2, 5, and/or 6.

**Equipment**

- cage assembly hardware: cage rods, cage plates, posts, post holders, pedestal mounts, table clamps, etc.
- two mirrors in kinematic mounts and associated hardware
- cage beam centering target(s)

**Final Setup**

The final setup you are working toward should eventually look like:

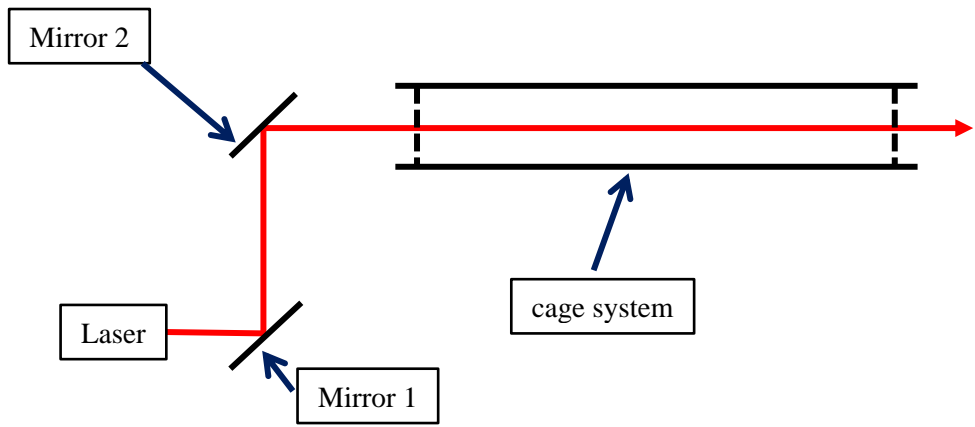

Carefully follow the outlined procedures to get there.

**Procedure**

**Part 1: Constructing the cage system and securing it to the table.**

30. Choose your cage system size (16 mm, 30 mm, or 60 mm) depending upon your available cage hardware and size of your optical components.  The sizes of 16 mm, 30 mm, or 60 mm corresponds to the distance between the cage rods.  Generally, 16 mm cage systems are used for 12.5 mm (or 0.5 inch) optics, 30 mm cage systems are used for 25 mm (or 1.0 inch) optics, and 60 mm cage systems are used for 50 mm (or 2.0 inch) optics.
31. Choose your cage system length.
32. Use a cage plate at each end of the cage rods to construct the cage system.

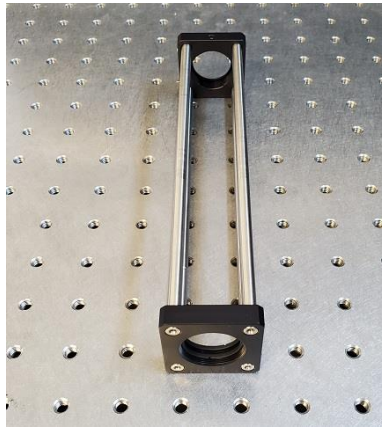

**Always have at least two mounting points for your cage system.**  Two mounting points (with as much distance between them as possible) make for a secure setup which will not easily misalign due to an inadvertent bump to the cage system.
*Try not to use a single mounting point* **for an entire cage system.** Single mounting point cage system clamps are prone to rotational misalignment about the single bolt if the cage system is bumped, especially for longer cage systems.

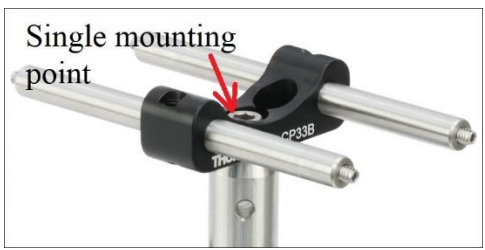
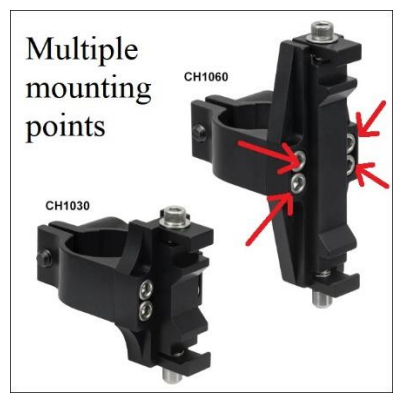

Cage mounting systems with single or multiple mounting points.

33. Choose two mounting points for your cage system. There are a few common ways to mount a cage system to the optics table.
One method is to use attach an optics post to the end cage plates using the threaded hole in the cage plate and mount it to the table using post holders and table clamps.

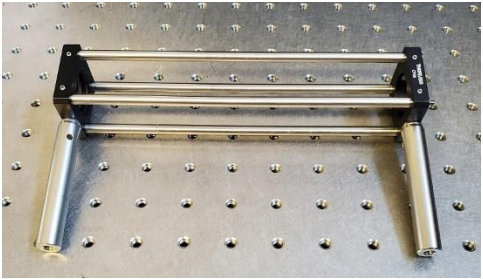  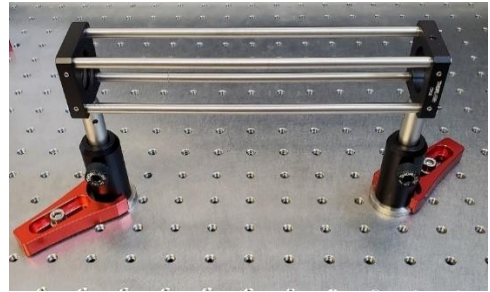

Posts attached to the cage plates.   Cage system using post holders and fork clamps.

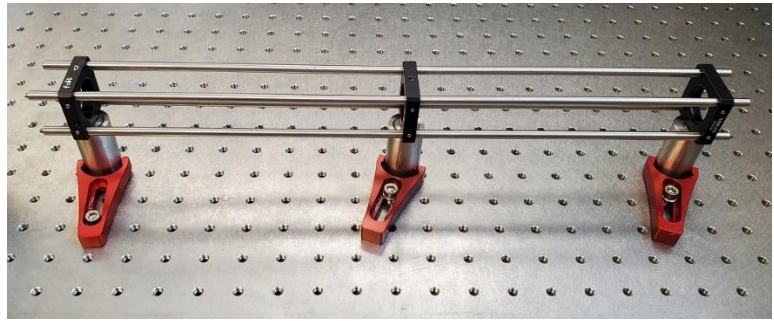

Cage system mounted using 1-inch pedestal posts and fork clamps.

Another method is to use a cage system mounting bracket attached to the top of an optics post. Be sure to use two of these.

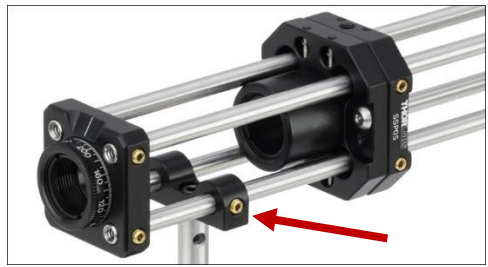

Cage mounting bracket on a post.

Be sure to use two of these, with one on each end of the cage system, for maximum stability.

Or, one could use a single cage system clamp for vertical posts.

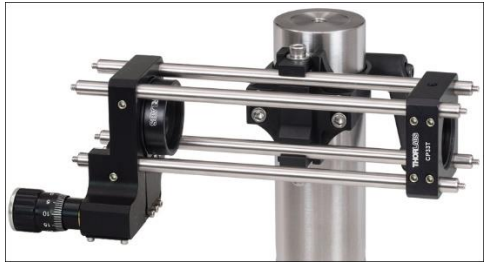

**Making the cage system parallel to the table bolt-hole lines and centered above a line.**

Before securing the cage system to the optics table carefully align the cage system to be parallel to the table bolt-hole lines.

34. Place the cage system on the table in the approximate location you want it to be.
35. Position your head such that you can look down over the cage rods with your line of sight perpendicular to the cage rods.  Choose a table bolt-hole line that is located at the approximate center of the cage system.  When your line of sight is along the chosen bolt-hole line then that bolt-hole line will appear to go straight away/towards you and will not appear to be at an angle, on converge to a point at a distance.  See Figure below.
36. Adjust the position of the cage system until the edge of the cage rods lines up exactly with the edge of a pair of bolt holes.  Choose two table bolt holes which are far apart and symmetric about your line of sight.

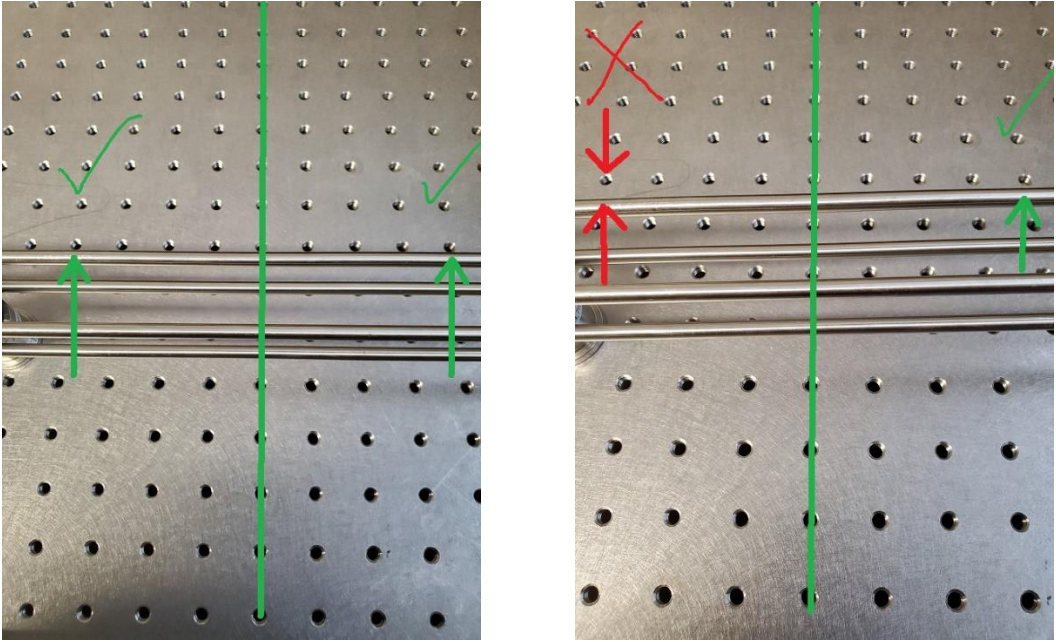

Center green line indicates your line of sight.  Note that the bolt-hole line of your line of sight appears to be straight towards/away from you and not at an angle.  Looking at the fourth hole from the center, the edge of the cage rod matches up with the edge of the bolt hole… if the cage system is parallel to the table bolt-hole lines.

37. Look directly down through the cage rods at your desired bolt-hole line for your desired beam path. Move the cage system such that the chosen bolt-hole line appears to be *exactly* centered between the cage rods.
38. Now look back along the perpendicular bolt-hole line (used in step 6 and 7) and double-check that the cage rods are still *exactly* parallel to the table's bolt-hole lines, as in step 7.
39. Secure the cage system mounts to the table using the appropriate table clamps.

**Center the beam through the cage system.**

Now that the cage system's location is centered along a table bolt-hole line and fixed into position, the overall approach here is to use previously mastered beam-walking skills (Tutorial 3: walk a beam horizontally and vertically) to walk the incident beam until it is perfectly centered through the cage system.

First, you will need a method to find locate the center of the cage system. There are several different pieces of hardware for finding the center of the cage system. Three common pieces of hardware are: (1) alignment plates, (2) threaded alignment targets, or (3) various home-made targets. The following figures are illustrations of some of these alignment tools.

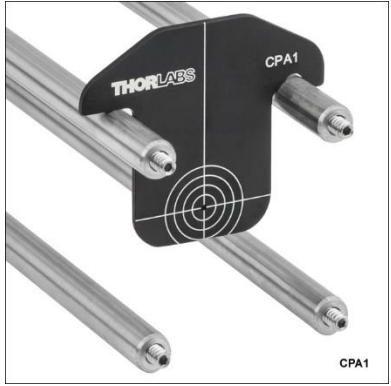 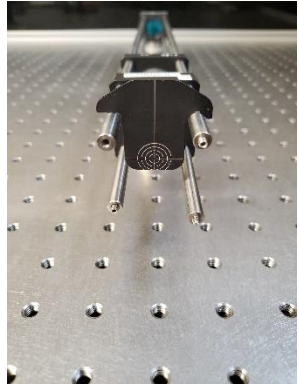

Cage alignment plate: which is quickly and easily removable and hangs from upper cage rods.

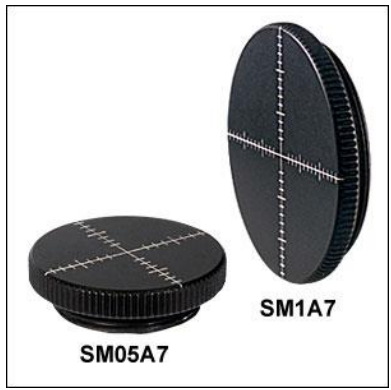 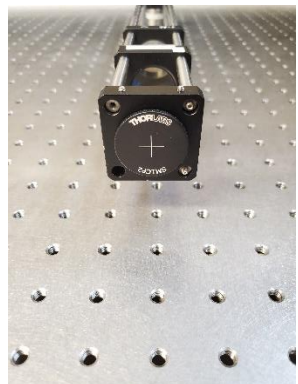

Threaded alignment target: which mounts to threaded cage plates. For using the near-near far-far alignment method place a cage plate at either end of the cage system for moving the target back and forth.

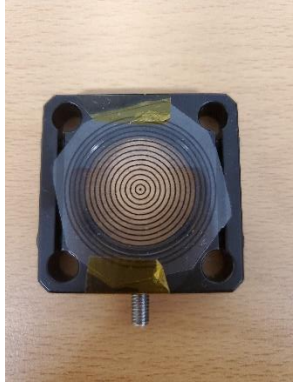 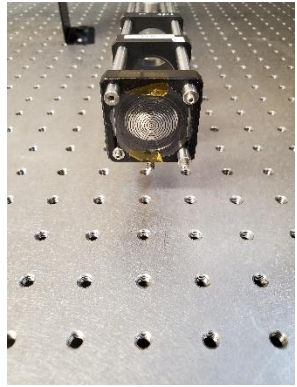

Home-made centering target on a cage plate. When assembling the cage system leave enough room on both ends to be able to move the centering plate back and forth during the alignment.

**Near-near far-far alignment**

Now we want to use the near-near far-far alignment method to perfectly center the beam through the cage system.

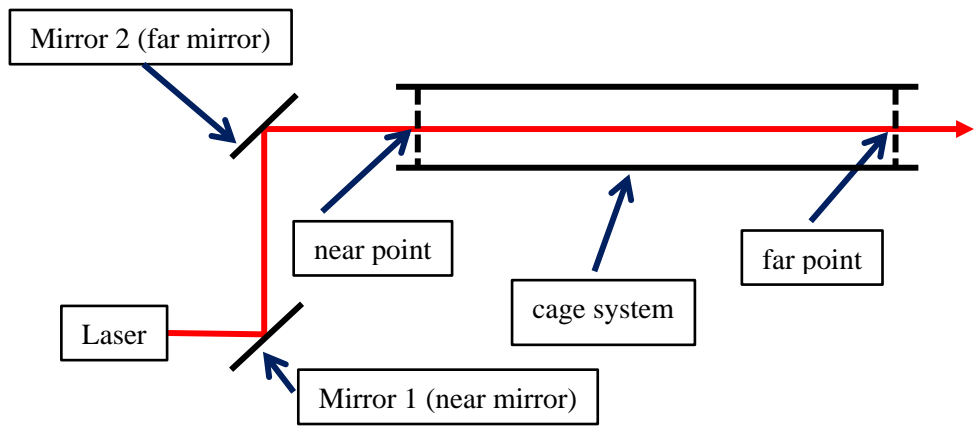

- Place your centering target at the near point in the cage system.
- Adjust the near mirror, Mirror 1, to center the beam at the near point target.
- Move the centering target to the far point in the cage system.
- Adjust the far mirror, Mirror 2, to center the beam at the far point target. (By adjusting this mirror the location of the beam at the near point has also slightly shifted.)
- Move the centering target back to the near point and recenter the beam using Mirror 1.
- Repeat this process of adjusting the near mirror to center the beam at the near target and adjusting the far mirror to center the beam at the far target until the beam is perfectly centered at both locations.

# Tutorial 8: align a lens to a beam path

**Initial Setup**

The initial setup for this skill should be the final setup for Tutorial 1, or Tutorial 2.

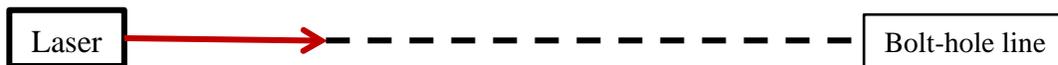

or

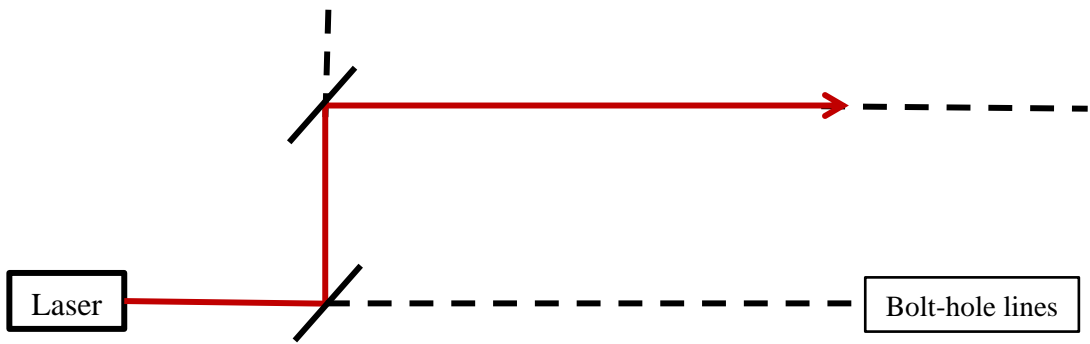

**Prerequisites**

There are no specific beam alignment prerequisites for this tutorial. This tutorial assumes that the laser path is already fixed and the user simply needs to insert a lens into the fixed beam path.

**Tutorial Goals**

The goals of this tutorial are to:

- understand the impact placing a lens into the path of a laser beam has on the change in the optical axis of the system

- learn about the proper equipment to use in order easily, and quickly, remove and replace a lens into the optical system
- learn to precisely adjust the horizontal and vertical positioning of a lens
- learn how to check if the lens is perfectly centered on the incident beam

**Equipment**

- Equipment necessary for Tutorial 1 or 2 shown above.
- Positive focal length lens, mount, post, post holder, table clamp.
- post clamp for lens (preferred), or nested mount system
- Viewing screen

**Final Setup**

The final setup you are working toward should eventually look like:

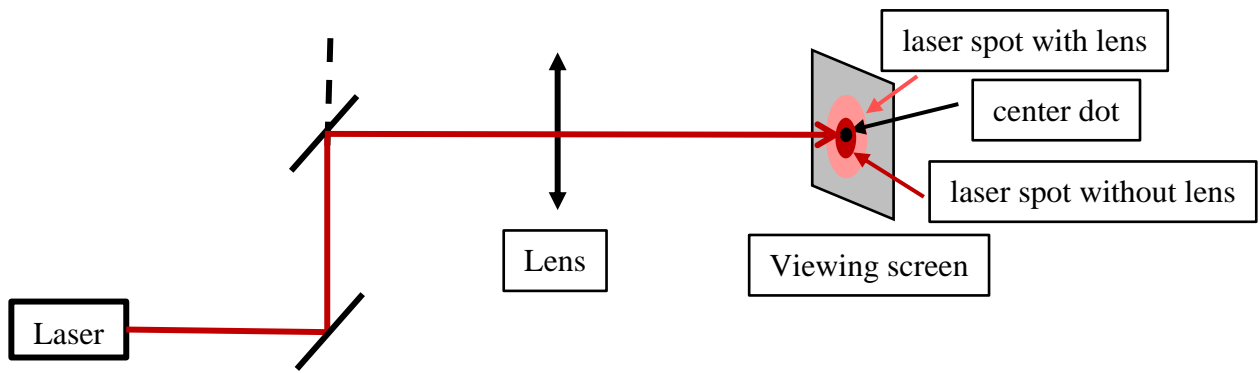

Carefully follow the outlined procedures to get there.

**Background Theory**

To understand the impact of placing a lens in the path of a freely propagating laser beam we will first use a simple ray diagram approach, and then a mathematical modeling approach.

## Ray Diagram Approach for a Single Ray

Using ray diagrams and a thin lens approximation, a ray that passes through the center of a lens continues along its incident path without any deviation. Figure 1 demonstrates this for a ray incident along the center axis of the lens.

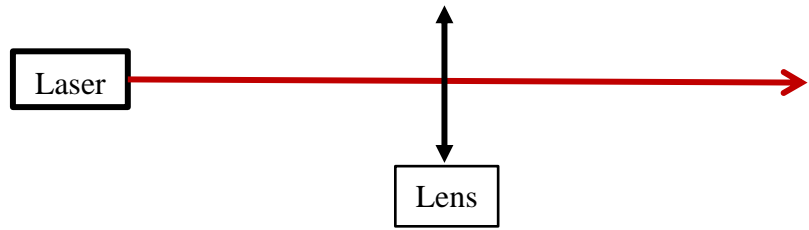

Fig. 1. A light ray incident upon the center of a lens along the lens' central axis.

If the light ray does not pass through the center of the lens its path will change. In other words, if the ray is incident upon the lens parallel to, but not co-linear with, the lens axis the ray will bend away from its initial direction. Figure 2 is an illustration of a laser beam incident upon a converging lens parallel to, but offset from, the lens axis.

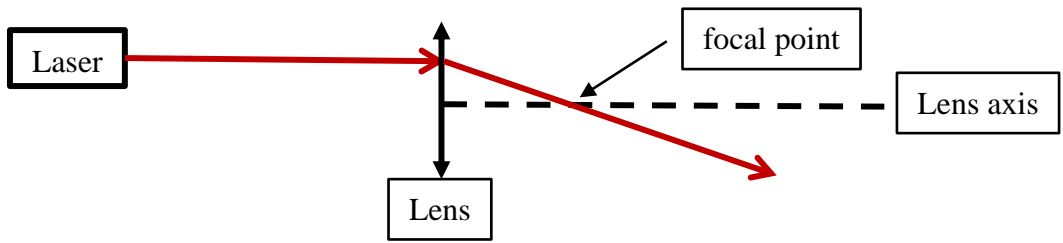

Fig. 2. A light ray incident parallel to, but offset from, the center of a lens.

Note in Fig. 2 that the path of the beam after the lens is *significantly different* than the path of the beam before the lens if the lens is not centered on the incident laser beam.

# Ray Diagram Approach for a Laser Beam with a Non-zero Width

All laser beams have a non-zero width. Using ray diagrams, the width of the laser beam can be simply illustrated using two rays for each "edge" of the laser beam. If the lens is centered on the incident laser beam, then the central axis of the laser is colinear with the axis of the lens, and the two rays representing the beam will be symmetrically located on either side of the lens's axis. As the rays progress beyond the lens they will *always* remain symmetric about the incident laser beam's central path. In other words, if the lens is perfectly centered on the beam:

- The beams with and without the lens in place **always have the same center**
- The effect of the lens on the beam is **only a change in size of the beam**

as illustrated in Figs. 3 and 4.

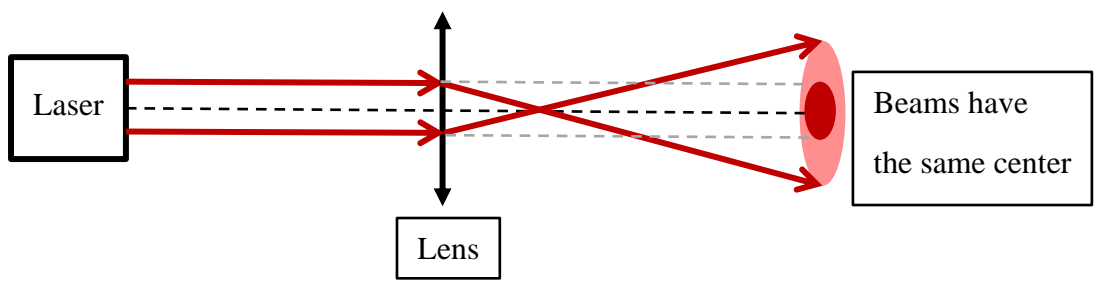

Fig. 3. When the lens is centered on the incident beam the beam with the lens has the same center as the beam without the lens in place. The darker red circle represents the beam without the lens in place, and the lighter red circle represents the beam with the lens in place. Note that the lens only changes the size of the beam not the location of the center of the beam.

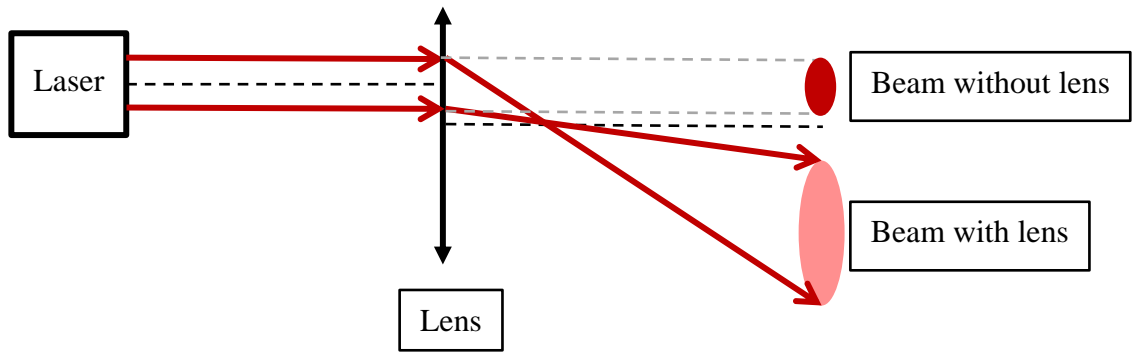

Fig.4. When the lens is not centered on the incident beam the lens changes both the size and center location of the beam after the lens.

## Mathematical Modeling Approach

The optical axis, in simple terms, is the center of the beam propagation direction. First, we will consider the simple case of a laser beam traveling through free space. To mathematically model the light propagation and the path of the laser beam we must first choose a coordinate system and its orientation. Using cylindrical coordinates, $(r, \theta, z)$, we would choose the center of the beam to be the *z*-axis and *r* would represent any distance away from the center of the beam. For this simple case, the *z*-axis would represent both the center of the beam and the optical axis of the system.

When light passes through a lens the optical axis *after* the lens is the center of the lens. If the center of the lens is not the same as the center of the laser beam before the lens, then the optical axis will shift according to the offset of the lens as illustrated in Fig. 5.

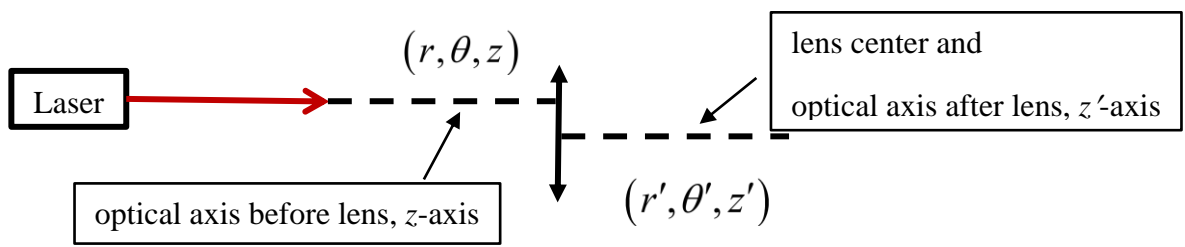

Fig. 5. The field of space before the lens has the $(r, \theta, z)$ coordinate system. The field of space after the lens uses the $(r', \theta', z')$ coordinate system. The lens position and orientation would be included in a coordinate transformation between the unprimed and primed systems.

If we were to mathematically model the complete beam path, we would need to use a different coordinate system after the lens of $(r', \theta', z')$ than before the lens. Mathematically modeling the complete beam path would involve first a determination of a coordinate transformation from $(r, \theta, z)$ to $(r', \theta', z')$ to represent the offset of the lens (and any possible differences in angle between $z$ and $z'$. Once the coordinate systems and the transformation functions between them is set, the user of the model needs to be meticulously careful of keeping track of the regions of validity of each coordinate system for each region when calculating the propagation of light fields to the point of interest. For optical systems of multiple lenses this can become quite cumbersome. All of this is to say that if the lens is not centered on the incident beam with the lens axis colinear to the incident optical axis then mathematically modeling the propagation of the light fields can become quite challenging.

However, the mathematical model simplifies *significantly* if a single coordinate system can be used for all regions of space. Regarding placing lenses in the beam path, a single uniform optical axis occurs only when the placement of each lens in the system is perfectly centered with its incident beam.

## Procedure

Overall, the procedure and setup for this skill is fairly easy. You just want to be able to quickly and easily remove and replace the lens to check that the beam with the lens in place has the same center as the beam without the lens in place.

**Establishing the center of the beam without the lens**

1. Note the focal length of the lens you want to insert into the beam path.
2. Note the location where you want to insert the lens.
3. Place a viewing screen in the beam path approximately 3 to 7 focal lengths beyond the desired location of the lens.
    a. If your viewing screen has a target pattern, center the pattern on the beam.

    b. If your viewing screen is simply a blank business card or piece of paper, then make a small dot with a pencil/pen to indicate the center of the beam.

**Hardware to allow quick and easy removal and replacement of the lens**

The key for checking your lens placement later on is to use optical hardware that allows the lens to be quickly removed and replaced from the optical setup without having to realign the lens each time.

One common acceptable method for having the lens easily removable is by using a post clamp so that the height of the lens is set, and the post holder and base is securely attached to the table to set the horizontal location of the lens, as illustrated in Fig. 6.

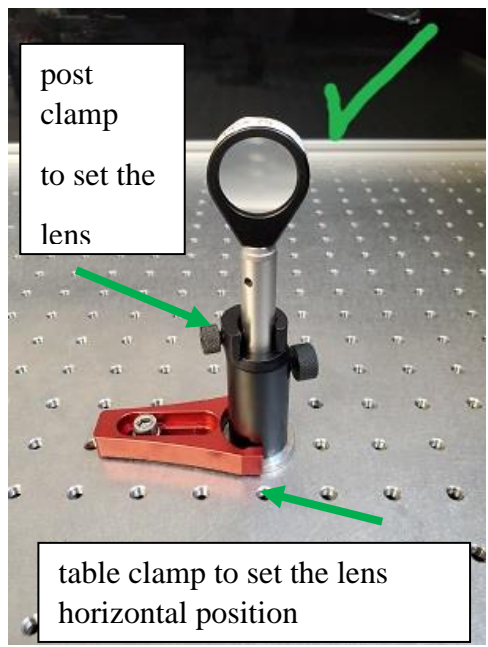

Fig. 6. Lens mounted in a fixed mount attached to a post. The table clamp on the postholder base sets the horizontal position of the lens. Using a post clamp (right) allows for the postholder thumb screw to be loosened and the post removed and replaced to exactly the same height.

Another hardware option to mount a lens such that it is easily removed and replaced is to use a nested mount system. A nested mount system is illustrated in Fig. 7. The mounting ring is attached to a post and remains a "permanent" part of the optical setup on the table. The lens is mounted in the optic carriage which slips into/out of the mounting ring and held in place via the thumbscrew on the top of the mounting ring.

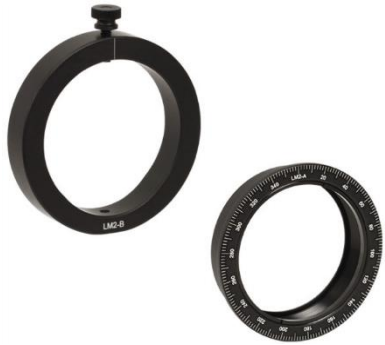

Fig. 7. Nested lens mount system.  The mounting ring (left) is attached to a post and remains in the optical system.  The lens is mounted in the optic carriage (right) which is easily removed and replaced into the mounting ring via the thumbscrew on the top of the mounting ring.

**Centering the lens**

4. Place the lens in the chosen mounting system into the beam path.  If you are using a post clamp, do not attach it to the post yet.
5. Observe the beam on the viewing screen. The beam should be noticeably larger than without the lens.  If not, move the viewing screen farther away from the lens until the beam is noticeably larger, remove the lens, and recenter the viewing screen on the laser.
6. Center the beam horizontally with the center of the target pattern (or directly above, or below, your pencil mark) by sliding the postholder and base on the optics table.
7. Tighten down the table clam to set the postholder and base into place.
8. Loosen the set screw in the postholder so that you can adjust the height of the post in the postholder.
9. Adjust the height of the beam on the viewing screen by sliding the post vertically in the postholder until the beam is vertically centered on the viewing screen indicator.
10. Tighten the postholder thumbscrew to hold the post in place.
11. Double-check the alignment of the beam on the viewing screen indicator as the lens might have moved slightly during the tightening process of either the table clamp or the postholder thumbscrew.
12. Now slip the post clamp over the post and tighten its set screw

**Double-checking the lens alignment**

13. With the lens in place check that the beam is centered on the viewing screen indicator.
14. Losen the thumbscrew on the postholder. The post should not move vertically due to the post clamp holding the post at the correct height.
15. Pull the lens/mount/post/post clamp assembly out of the postholder.
16. Check that the beam is still incident on the same viewing screen indicator now that the lens is removed from the system. If not, go back to step 6.
17. Replace the lens/mount/post/post clamp assembly back into the postholder. The post clamp should be forcing the lens to be located at the pre-set height.
18. Check one last time that the beam is still centered on the viewing screen indicator with the lens in the beam path.

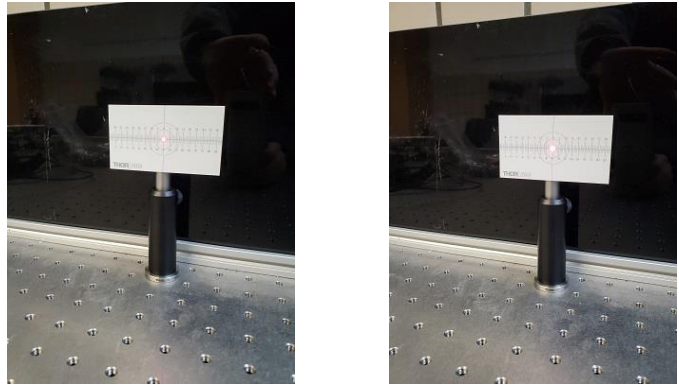

Fig. 8. (left) Beam centered on the viewing screen without the lens in place, and (right) beam centered on the viewing screen with the lens in place.

# How to Secure Optical Hardware to an Optics Table

Securing optical hardware to an optics table, and not causing long-term use damage to the equipment, may not be as obvious as it seems.

## Bolts and Washers

*Always* use a washer under a bolt head when securing anything with a bolt.

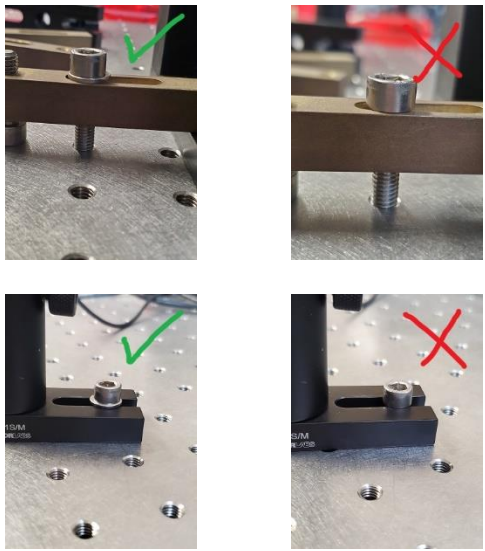

Damage occurs to the optical hardware when a washer is not used. The bolt is made of steel and optical hardware is typically made of softer materials such as aluminum or brass. When the bottom of the steel bolt head comes into contact with the softer aluminum of the hardware it will scratch the hardware surface. If the bolt is tightened down (as it should be nice and tight) then the bottom of the bolt will dig into, or gouge, the optical hardware causing permanent damage.

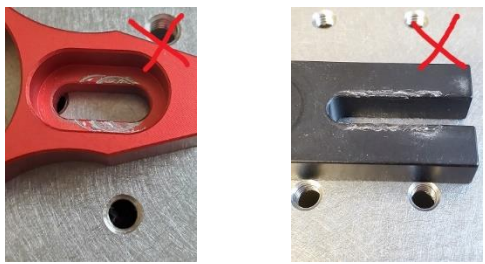

Other than permanently damaging the equipment, the grooves created by tightening a bolt without a washer can create a frustrating alignment experience. If the surface of the optical hardware is grooved instead of smooth what can happen is that the next time it is tightened down the bolt will shift the position of the hardware so that it settles down into a groove as it is tightened.  This can be frustrating because you may have just spent time getting the hardware into just the right position so that your beam alignment is perfect only to find that the optical hardware shifted when you secure it to the table.

## To Use a Clamp or Not to Use a Clamp?

There are essentially two different possibilities to secure a piece of optical hardware to an optics table: (1) use a bolt with a washer through the slots, or holes, provided in the optical hardware, or (2) use a table clamp.

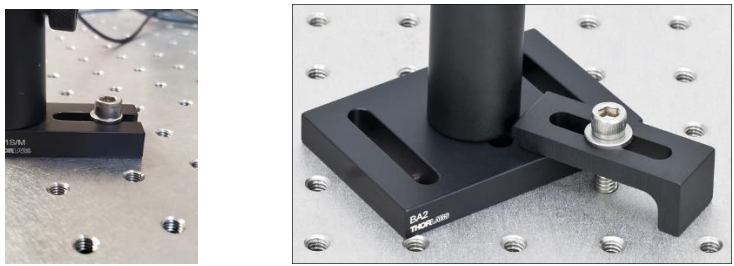

Which method you use is up to you.  There is an advantage to using table clamps.  When you use the slots, or holes, provided in the optical hardware you are limited in the placement, and rotation, of the optic for the slots, or holes, to line up with a table bolt hole.  When you use a table clamp you are not limited in the placement of the optic.  You have the freedom to place and rotate the optical component to have any position you need.  Once you have the optical hardware *exactly* where you want it the table clamp has the freedom to be moved around in order to line up with a table bolt hole.

## Some Common Types of Table Clamps

### General Purpose Table Clamp

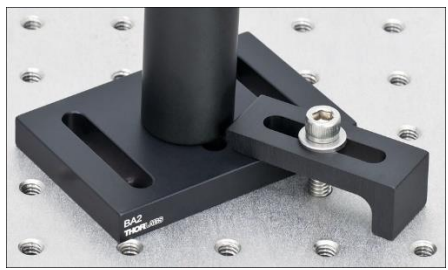

General purpose table clamps are useful for holding down optical hardware mounted to standard bases with a height of 3/8", or 10 mm.

**Clamping Fork**

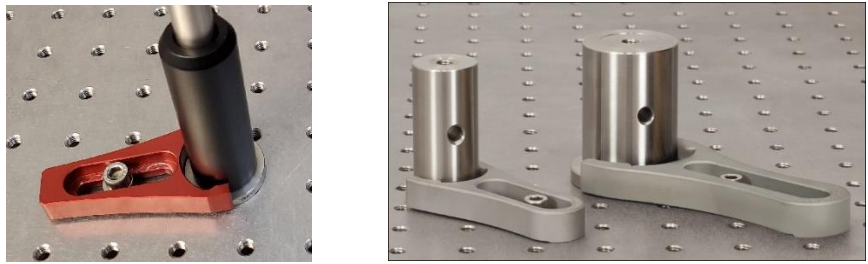

Clamping forks are used with optics with a pedestal base. Pedestal bases have a smaller footprint than standard slotted bases.

**Adjustable Height Table Clamps**

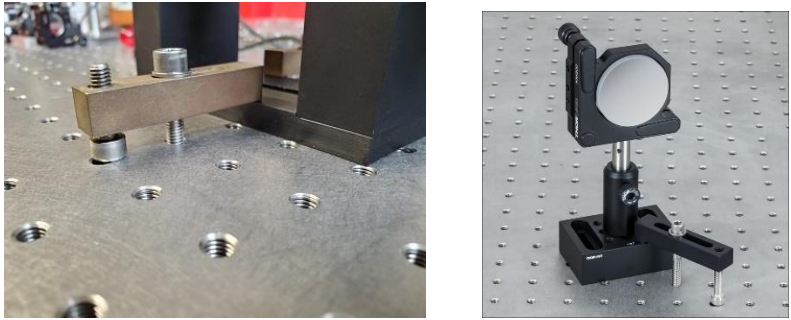

Adjustable height table clamps use two bolts: (1) one bolt to adjust the height of the clamp, and (2) one bolt to secure the hardware to the table.

To prevent damage to the clamp or to the table:

- The height-adjust bolt *must be* used "upside down"
- The securing bolt *must have* a washer under the bolt head
- Never adjust the height of the clamp with the securing bolt in place

The height-adjust bolt is placed such that the head of the bolt is in contact with the optics table. This is to prevent damage to the optics table. The head of a bolt is a relatively smooth surface. The threaded end of a bolt will have burs or sharp edges which can damage the surface of the optics table.

To adjust the height:

- remove the securing bolt
- pick the adjustable table clamp up in your hands and spin the adjustment bolt with your fingers

- place the clamp back on the table and optical hardware and check the height of the clamp
- adjust the position of the adjustment bolt such that the clamp is roughly parallel to the table surface
- Once the height is set then insert the securing bolt and tighten

# Proficiency Test 1: Laser output along a table bolt hole line

**Task**

*Safely* setup a laser such that the beam path *exactly* follows a chosen bolt hole line along the table.

**Note:**

Your instructor can stop this activity at any time if proper laser safety procedures are not being followed.

# Proficiency Test 2: Horizontal shift along parallel table bolt-hole lines

**Setup**

The initial setup for this skill should be the final setup for **Tutorial 1: laser output along a table bolt-hole line**.

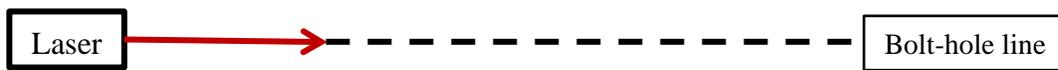

**Task**

*Safely* construct a setup to shift the laser path horizontally such that final laser beam *exactly* follows a parallel bolt hole line along the table with a 15-20 cm offset from the beam path exiting the laser.

**Notes**

- beam path changes must be *exactly* 90°
- when the beam changes directions it must do so *exactly* centered above a table bolt hole
- Your instructor can stop this activity at any time if proper laser safety procedures are not being followed.

# Proficiency Test 3: Walk a beam horizontally and vertically

**Setup**

The initial setup for this skill should be the final setup for **Tutorial 2: horizontal shift along parallel table bolt-hole lines**.

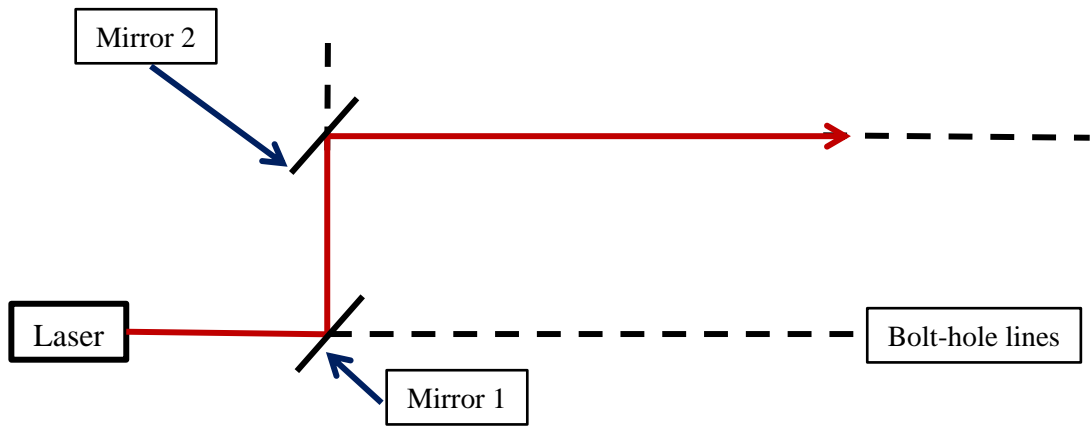

**Task 1**

Center an iris on the beam path after Mirror 2 near the mirror (about 12.5 to 15 cm). Center another iris on the beam path far (about 30 to 50 cm) from Mirror 2.

Have your instructor check this task.

**Task 2**

Your instructor just moved the irises both horizontally and vertically.

Use the new locations of the irises as the new desired beam path. Use the horizontal and vertical beam walking method to recenter the beam on the irises.

**Notes:**

- Your instructor can stop this activity at any time if proper laser safety procedures are not being followed.

# Proficiency Test 4: Align a second laser beam along an established beam path

**Initial Setup**

The initial setup for this skill should be the final setup for **Tutorial 3: Walk a beam horizontally and vertically**.

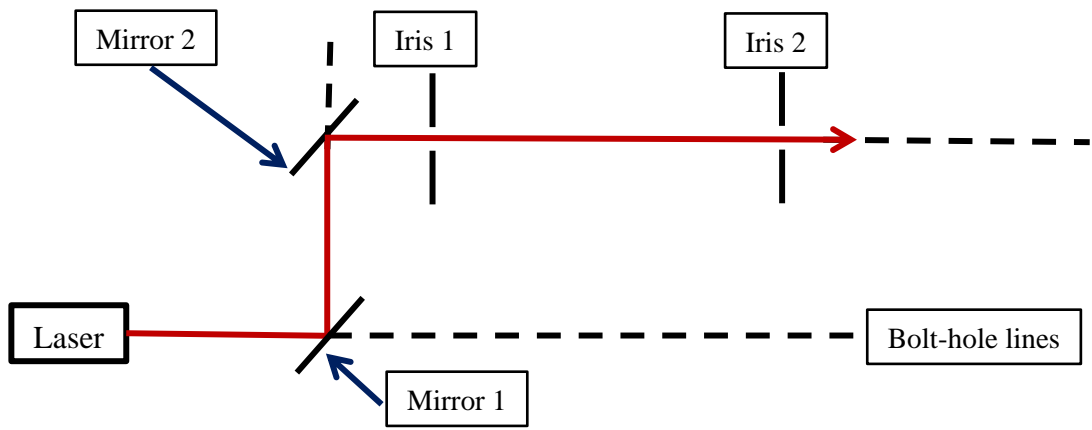

**Task**

Align a second laser beam such that it follows the *exact* same path as the original, *and* you can alternate between the two laser beams on the final path by use of a flipper mirror.

**Notes:**

- Your instructor can stop this activity at any time if proper laser safety procedures are not being followed.

# Proficiency Test 5: Vertical shift without polarization rotation

**Setup**

The initial setup for this skill should be the final setup for **Tutorial 1: laser output along a table bolt-hole line.**

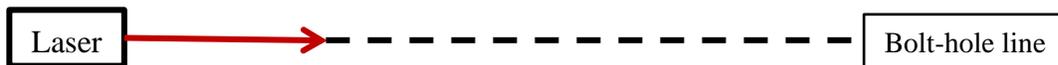

**Task**

*Safely* construct a setup to shift the laser path vertically such that final laser beam *exactly* follows the same bolt-hole line of the beam path exiting the laser.

**Notes:**

- beam path changes must be *exactly* 90°
- when the beam changes directions it must do so *exactly* centered above a table bolt hole
- Your instructor can stop this activity at any time if proper laser safety procedures are not being followed.

# Proficiency Test 6: Vertical shift with a 90° polarization rotation

**Setup**

The initial setup for this skill should be the final setup for **Tutorial 1: laser output along a table bolt-hole line.**

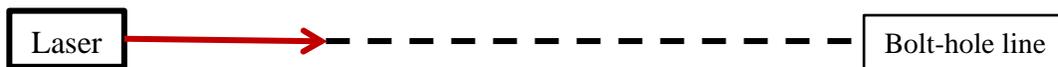

**Task**

*Safely* construct a setup to shift the laser path vertically such that final laser beam:

- is shifted vertically by about 7.5 to 15 cm,
- is parallel to the surface of the table,
- *exactly* follows a bolt-hole line of the table, and
- has a polarization rotation of *exactly* 90°.

You must also demonstrate that you can measure the polarization of the beam and verify that it has been rotated by 90°.

**Notes:**

- beam path changes must be *exactly* 90°
- when the beam changes directions it must do so *exactly* centered above a table bolt hole
- Your instructor can stop this activity at any time if proper laser safety procedures are not being followed.

# Proficiency Test 7: Align a beam through a cage system

**Setup**

The initial setup for this skill should be the final setup for Tutorials 2, 5, or 6.

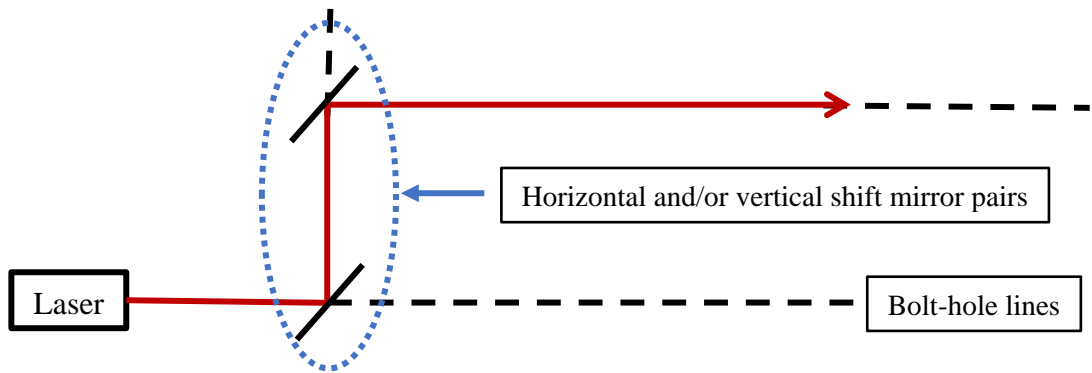

**Task**

*Safely* construct a setup so that a laser beam passes through the cage system such that:

- the cage system is properly mounted to the optics table, *and*
- the cage system is exactly parallel to the table bolt-hole lines, *and*
- the cage system is exactly centered above the chosen bolt-hole line, *and*
- the laser beam passes exactly through the center of the cage system.

**Notes:**

- Your instructor can stop this activity at any time if proper laser safety procedures are not being followed.

# Proficiency Test 8: Align a lens to a beam path

**Setup**

The initial setup for this skill should be the final setup for Tutorial 1, or Tutorial 2.

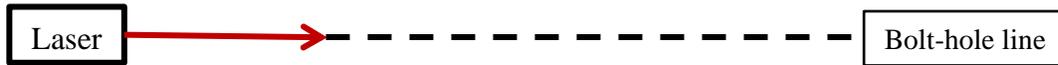

or

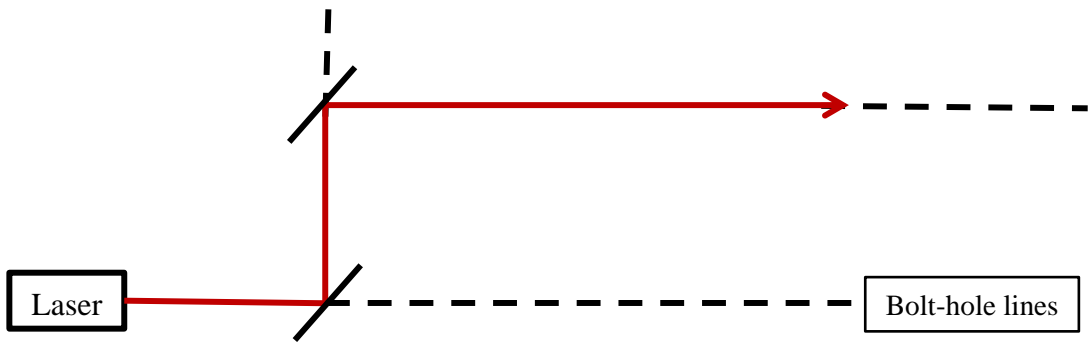

**Task**

*Safely* construct a setup so that a lens in placed *exactly* centered on the beam.

- the setup should allow the lens to be quickly and easily removed, *and*
- the setup should allow the lens to be quickly and easily replaced in the *exact* location.

**Notes:**

- Your instructor can stop this activity at any time if proper laser safety procedures are not being followed.

# Instructor's Guide 1: Laser output along a table bolt hole line

**Initial Setup**

There is no prior setup necessary for this activity. The student should be starting with an empty optics table.

**Equipment**

- laser
- ability to clamp the laser to the table
    - some laser housings allow for them to be set on the table with edges available for table clamps
    - some laser housings have an M6 thread on the bottom
        - post with an M6 threaded rod
        - post holder
        - base
        - table clamps and bolts
- Wooden ruler, ~30 cm
- Beam block

**Final Setup**

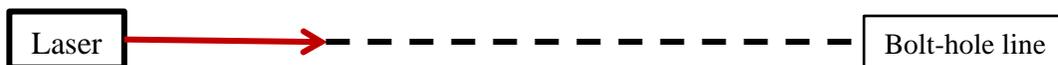

# Evaluation Checklist

|   | |
|---|---|
|   | **Laser Safety**<br>• beam terminates on beam block at all times while the student is working<br>• student never bends down towards, or into, the laser beam path |
|   | **Near bolt hole alignment**<br>• beam is cut in half by the edge of the ruler when the lower edge of the ruler is held flush against the table and the corner is centered on the far side of the bolt hole.<br>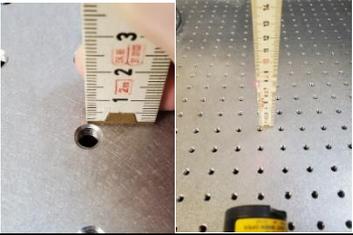 |
|   | **Far bolt hole alignment**<br>• beam is cut in half by the edge of the ruler when the ruler edge is centered on the far side of the far bolt hole |
|   | **Laser secured to the table**<br>• Table clamp is used properly<br>• all bolts on optical hardware have washers to prevent damage to hardware |

# Instructor's Guide 2: Horizontal shift along parallel table bolt-hole lines

**Initial Setup**

The initial setup for this skill should be the final setup for **Tutorial 1: laser output along a table bolt-hole line**.

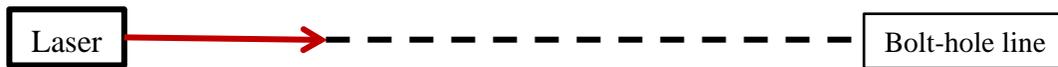

**Equipment**

- laser
- ability to clamp the laser to the table
    - some laser housings allow for them to be set on the table with edges available for table clamps
    - laser housing with an M6 thread on the bottom
        - M6 threaded rod
        - post
        - post holder
        - base
        - table clamps and bolts
- Wooden ruler, ~30 cm
- Beam block
- Two mirrors on kinematic mounts, associated posts, post holders, bases, and table clamps

**Final Setup**

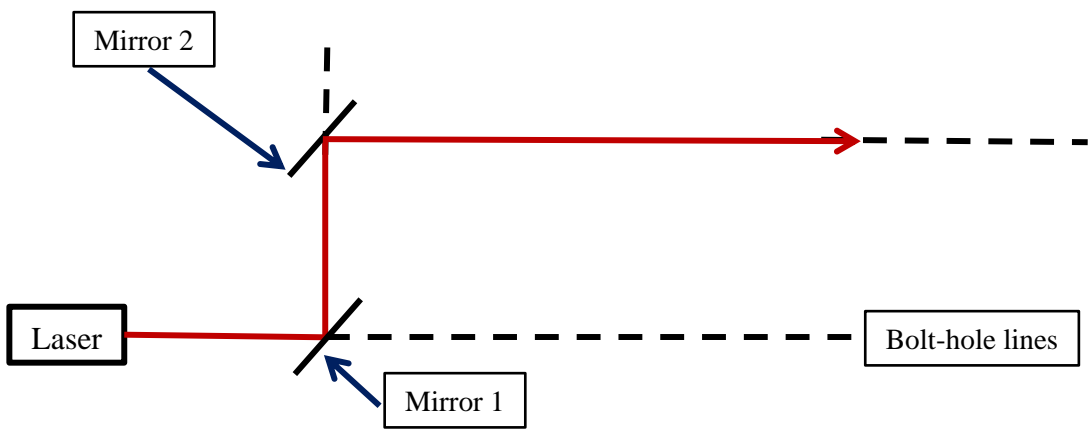

## Evaluation Checklist

| | |
|---|---|
| | **Handling of the mirrors**<br>• Did they not touch the surface of the mirror with their bare fingers |
| | **Near bolt hole alignment**<br>• beam is cut in half by the edge of the ruler when the ruler edge is centered on the far side of a bolt hole just after Mirror 2. |
| | **Far bolt hole alignment**<br>• beam is cut in half by the edge of the ruler when the ruler edge is centered on the far side of a bolt hole a good distance (at least 50-70 cm) from Mirror 2. |
| | **Beam is parallel to the table**<br>• Laser spot is the same height on the ruler at both the near and far positions |
| | **Laser and optical hardware properly secured to the table**<br>• Table clamps are used properly, all bolts have washers |

# Instructor's Guide 3: Walk a beam horizontally and vertically

**Initial Setup**

The initial setup for this skill should be the final setup for **Tutorial 2: horizontal shift along parallel table bolt-hole lines**.

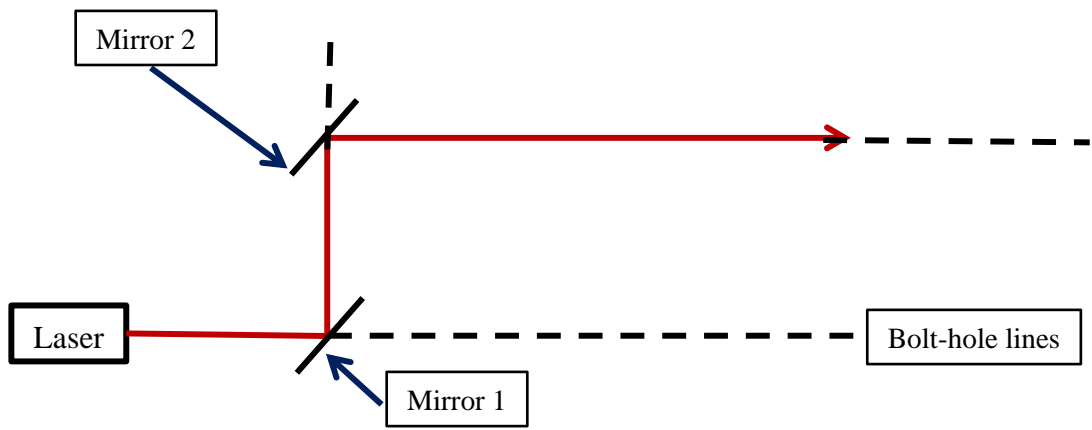

**Equipment**

- Equipment necessary for Tutorial 2
- Two irises and associated posts, post holders, bases, and table clamps

**Final Setup**

The final setup should look like:

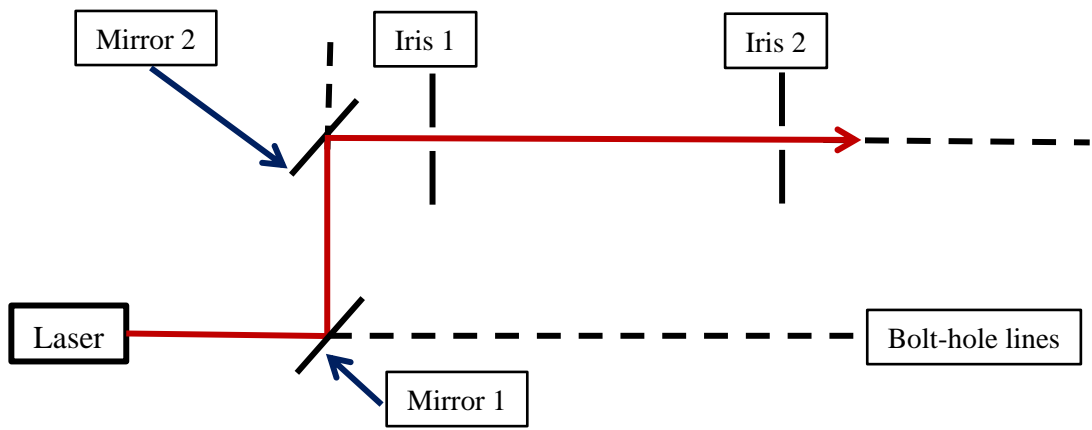

## Evaluation Checklist, Part 1: initial beam path

|   | |
|---|---|
|   | **Near iris alignment before walking the beam**<br>• Beam passes through the center of the near iris when closed. Check by observing the beam after the iris while opening and closing the iris. The iris shadow should open and close *concentrically* on the beam. |
|   | **Far iris alignment before walking the beam**<br>• Beam passes through the center of the far iris when closed. |
|   | **Optical hardware properly secured to the table**<br>• Table clamps are used properly, all bolts on optical hardware have washers |

**Iris Movement by the Instructor**

After the student has successfully centered each iris on the beam and using the check list above, the instructor should:

- Slide each iris horizontally (in the same direction) about 3-4 mm
- Change the height of each iris (in the same direction) about 3-4 mm
- Ask the student to use the beam walking method to recenter the beam through each iris

## Evaluation Checklist, Part 2: after walking the beam

|   | |
|---|---|
|   | **Near iris alignment after walking the beam**<br>• Beam passes through the center of the near iris when closed |
|   | **Far iris alignment after walking the beam**<br>• Beam passes through the center of the far iris when closed |

# Instructor's Guide 4: Align a second laser beam along an established beam path

**Initial Setup**

The initial setup for this skill should be the final setup for **Tutorial 3: Walk a beam horizontally and vertically**.

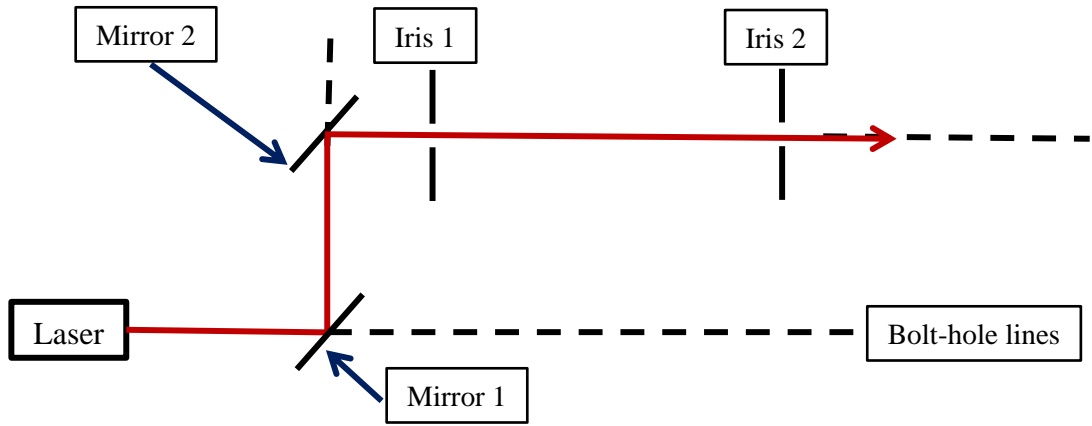

**Equipment**

- Equipment necessary for Tutorial 2 or 3 shown above
- Second laser
- Mirror, mount, post, post holder, table clamp
- Flipper mirror and mounting hardware

**Final Setup**

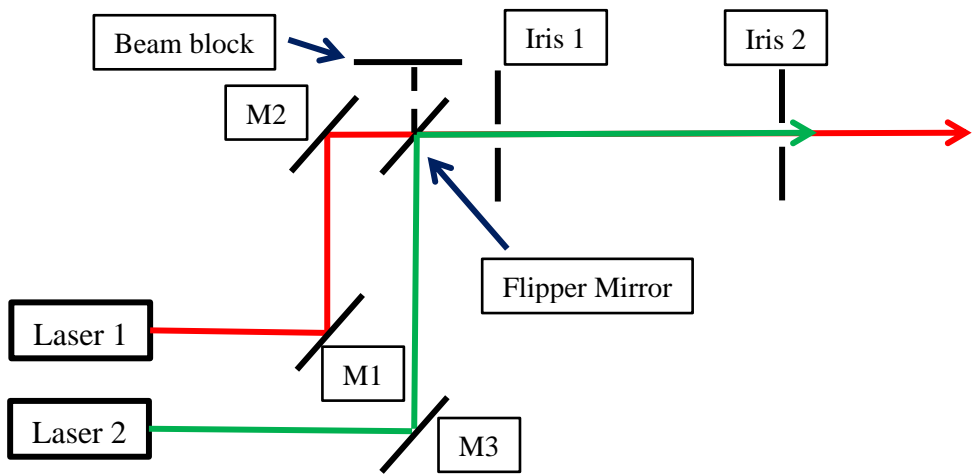

## Evaluation Checklist

|   |   |
|---|---|
|   | **Laser 2 beam block**<br>• Laser 2 safely terminates on a beam block when the flipper mirror is in the down position |
|   | **Near iris alignment**<br>• With the flipper mirror down Beam 1 passes through the center of the near iris when closed. Check by observing the beam after the iris while opening and closing the iris. The iris shadow should open and close *concentrically* on the beam.<br>• With the flipper mirror up Beam 2 passes through the center of the near iris |
|   | **Far iris alignment**<br>• Beam 1 passes through the center of the far iris when closed.<br>• Beam 2 passes through the center of the far iris when closed. |
|   | **Handling of the mirrors**<br>• Did they not touch the surface of the mirror with their bare fingers |
|   | **Optical hardware properly secured to the table**<br>• Table clamps are used properly, all bolts on optical hardware have washers |

# Instructor's Guide 5: Vertical shift without polarization rotation

**Setup**

The initial setup for this skill should be the final setup for **Tutorial 1: laser output along a table bolt-hole line.**

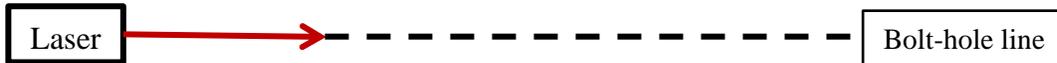

**Equipment**

- laser
- ability to clamp the laser to the table
    - some laser housings allow for them to be set on the table with edges available for table clamps
    - laser housing with an M6 thread on the bottom
        - M6 threaded rod
        - post
        - post holder
        - base
        - table clamps and bolts
- Wooden ruler, ~30 cm
- Beam block
- Two mirrors on kinematic mounts, associated posts, post holders, bases, and table clamps

**Final Setup**

The final setup should be a pair of mirrors aligned vertically with the final beam output parallel to the incident beam.

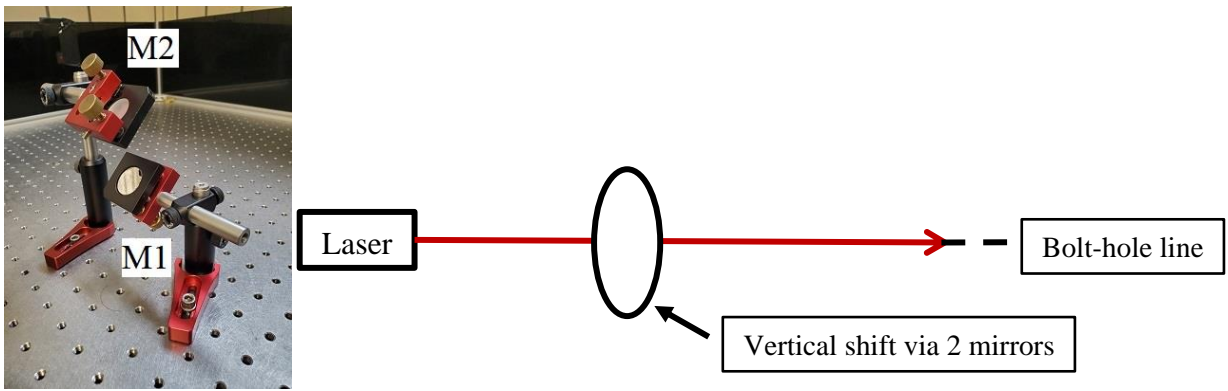

## Evaluation Checklist

| | |
|---|---|
| | **Near bolt hole alignment**<br>• beam is cut in half by the edge of the ruler when the ruler edge is centered on the far side of a bolt hole just after Mirror 2. |
| | **Far bolt hole alignment**<br>• beam is cut in half by the edge of the ruler when the ruler edge is centered on the far side of a bolt hole a good distance (at least 50-70 cm) from Mirror 2. |
| | **Beam is parallel to the table**<br>• Laser spot is the same height on the ruler at both the near and far positions |
| | **Handling of the mirrors**<br>• Did they not touch the surface of the mirror with their bare fingers |
| | **Laser and optical hardware properly secured to the table**<br>• Table clamps are used properly, all bolts have washers |

# Instructor's Guide 6: Vertical shift with a 90° polarization rotation

**Setup**

The initial setup for this skill should be the final setup for for **Tutorial 1: laser output along a table bolt-hole line.**

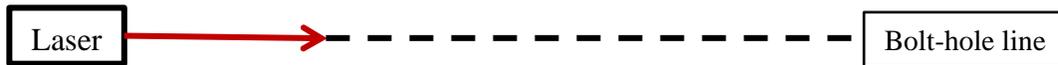

**Prerequisites**

The prerequisite skills for this tutorial are to be able to mount mirror into the beam path such that the reflected beam is *exactly* perpendicular to the incident beam, and to be able to steer the resultant beam both horizontally and vertically onto a desired final beam path. Tutorials for these skills are in:

- Tutorial 2: horizontal shift along parallel table bolt-hole lines
- Tutorial 3: walk a beam horizontally and vertically

**Equipment**

- laser clamped to the table
- Wooden ruler, ~30 cm
- Beam block
- Two mirrors on kinematic mounts, associated posts, post holders, bases, and table clamps
- Linear polarizer mounted in a rotation mount with an engraved scale
- Laser power meter

**Final setup**

The final setup should be a pair of mirrors aligned vertically with the final beam output perpendicular to the incident beam.

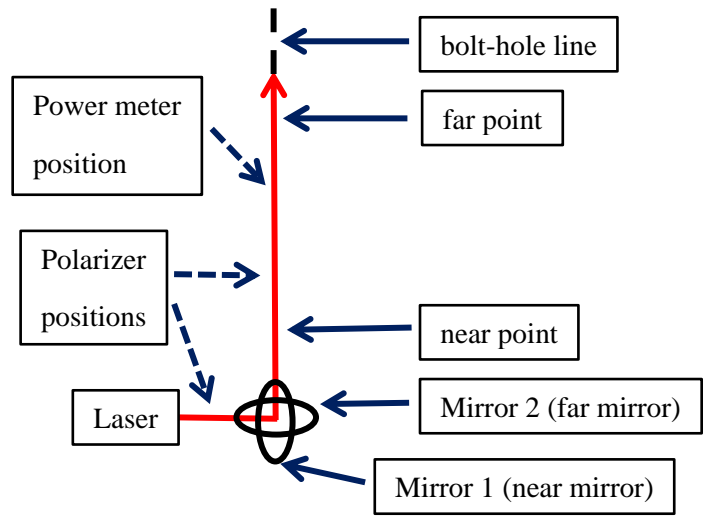

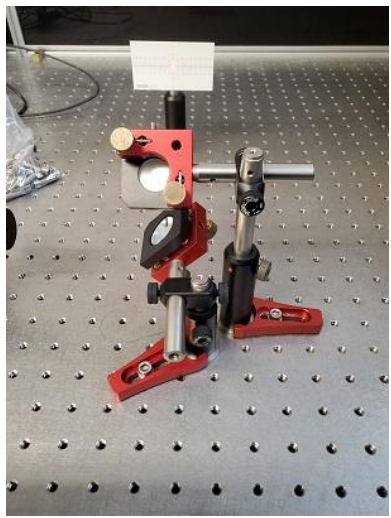
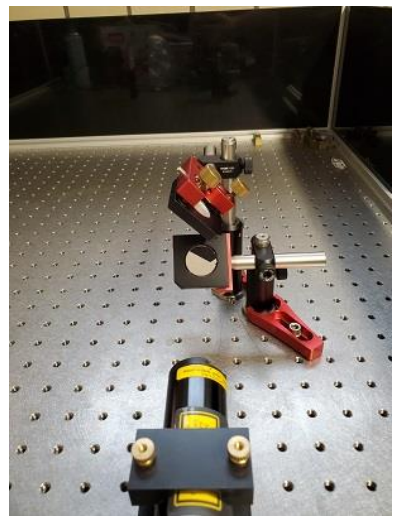

**Mirrors are centered above holes in the table**

You can check if the center of a mirror is *exactly* centered over a particular bolt hole by looking *directly* down at the mirror/bolt hole from above.

**How do you know if you are looking *directly down* at the chosen bolt hole?**

You cannot see the hole directly under the mirror, but you can see the two bolt-hole lines that intersect at the chosen hole, and sets of holes symmetrically around these lines.

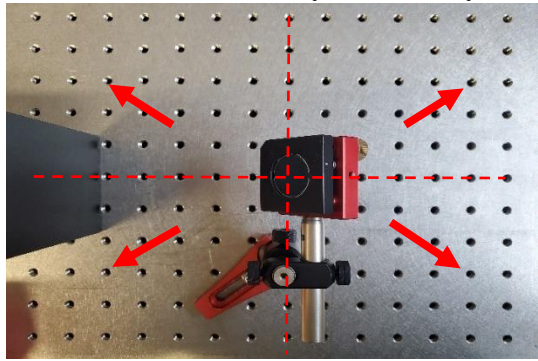

Pick a few holes that are a symmetrical distance away from directly under the mirror. Move your head around such that you are looking straight down at the mirror *and* the angles of the thread patterns within the symmetrically located holes are also symmetrical.

## Evaluation Checklist

| | |
|---|---|
| | **Student measured the polarization before and after the shift** <br> • Observe the student measuring the polarization before and after the shift with a polarizer and a power meter |
| | **Polarization shift is exactly 90°** <br> • Measure the polarization before and after the shift with a linear polarizer mounted in a rotation mount with an engraved scale and verify that the shift is 90°. |
| | **Mirror centers are directly above a bolt hole in the table** |
| | **Near bolt hole alignment** <br> • beam is cut in half by the edge of the ruler when the ruler edge is centered on the far side of a bolt hole just after Mirror 2. |
| | **Far bolt hole alignment** <br> • beam is cut in half by the edge of the ruler when the ruler edge is centered on the far side of a bolt hole a good distance (at least 50-70 cm) from Mirror 2. |
| | **Beam is parallel to the table** <br> • Laser spot is the same height on the ruler at both the near and far positions |
| | **Handling of the mirrors** <br> • Did they not touch the surface of the mirror with their bare fingers |
| | **Laser and optical hardware properly secured to the table** <br> • Table clamps are used properly, all bolts have washers |

# Instructor's Guide 7: Align a beam through a cage system

**Setup**

The initial setup for this skill should be the final setup for Tutorials 1, 5, or 6.

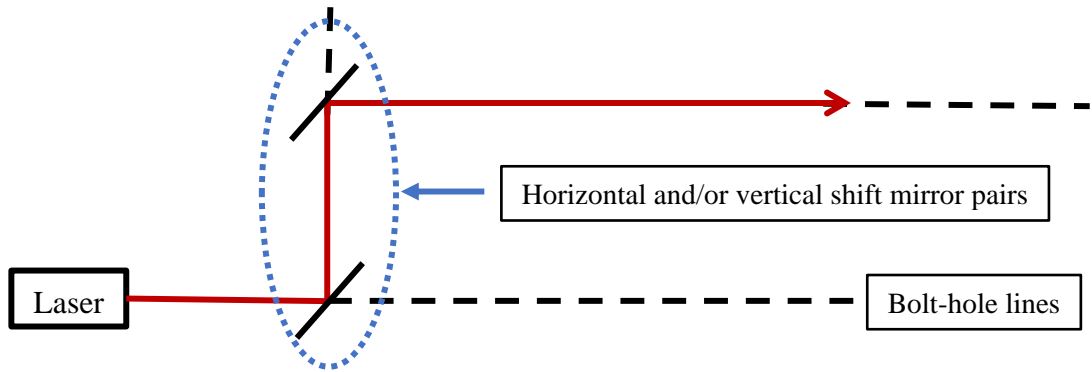

**Equipment**

- Equipment necessary for Tutorial 1, 5, or 6 shown above.
- cage assembly hardware: cage rods, cage plates, posts, post holders, pedestal mounts, table clamps, etc.
- two mirrors in kinematic mounts and associated hardware
- cage beam centering target(s)

**Final Setup**

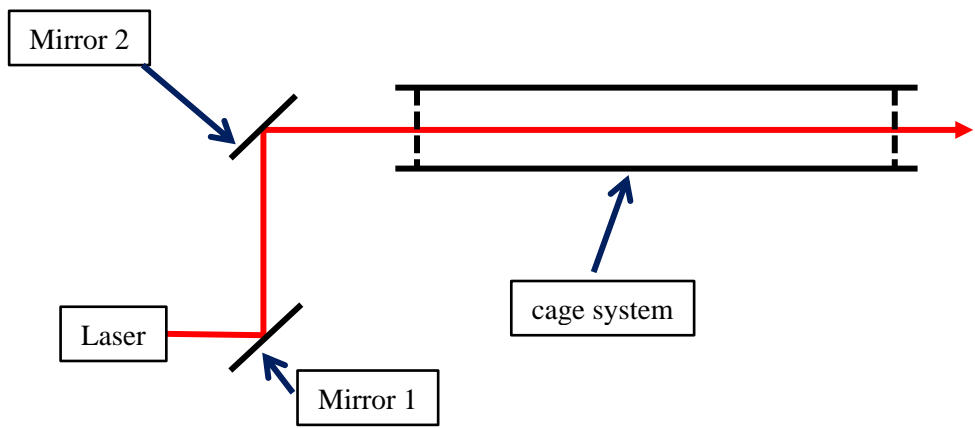

# Evaluation Checklist

| | |
|---|---|
| | **Cage system has more than one mounting point**<br>• a single mounting point is prone to rotational misalignment if bumped |
| | **Cage system is parallel to table bolt-hole lines**<br>• view from above along perpendicular center line<br>• use symmetrically located bolt holes and adjust your head location such that the edge of the holes line up with the edge of the cage system<br>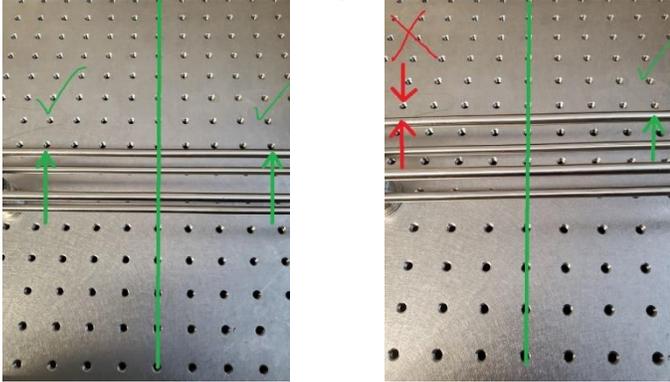 |
| | **Cage system is properly secured to the table using table clamps** |
| | **The beam is perfectly centered on cage center target at the near point** |
| | **The beam is perfectly centered on cage center target at the far point** |
| | **All optical hardware properly secured to the table**<br>• Table clamps are used properly, all bolts on optical hardware have washers |

# Instructor's Guide 8: Align a lens to a beam path

**Setup**

The initial setup for this skill should be the final setup for Tutorial 1, or Tutorial 2.

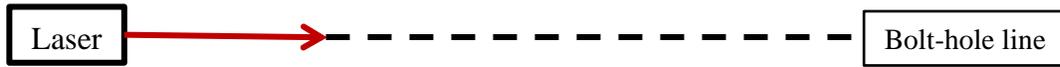

or

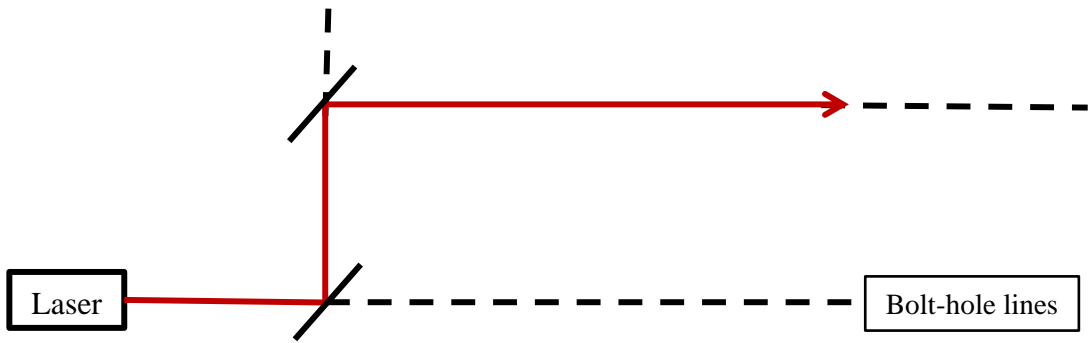

**Equipment**

- Equipment necessary for Tutorial 1 or 2 shown above.
- Positive focal length lens, mount, post, post holder, table clamp.
- post clamp for lens (preferred), or nested mount
- Viewing screen

**Final Setup**

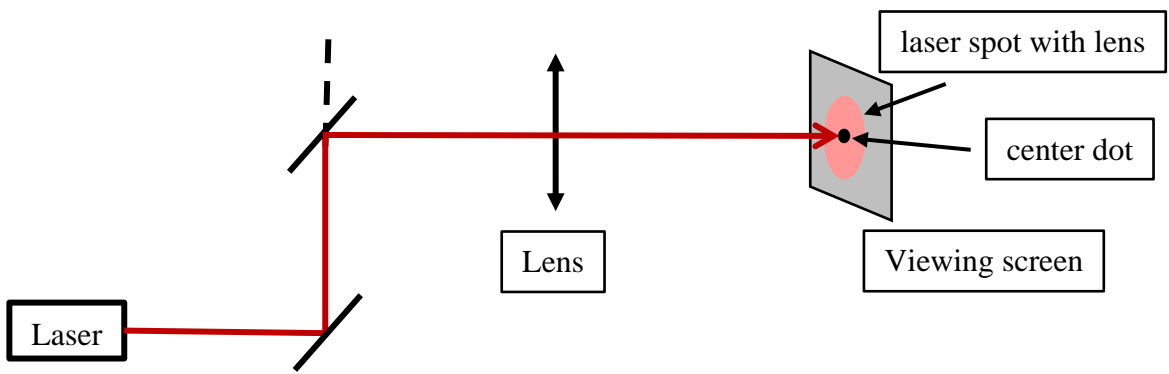

The final student setup as a minimum would just have the lens in the beam path and the beam terminating on the viewing screen. The student might have used the iris for alignment and left that in the setup, which would be OK.

**Evaluation Procedure**

Is the lens easily removable from the system? If not, tell the student to start over, and/or refer them to the tutorial.

The most common acceptable method for having the lens easily removable is by **using a post clamp** so that the height of the lens is set, and the post holder and base is securely attached to the table to set the horizontal location of the lens, as illustrated in the first to pictures below. Another method is to **use nested mounts** where the outer ring stays in place and the lens is mounted to the inner mount which can easily be removed and replaced, as illustrated on the right.

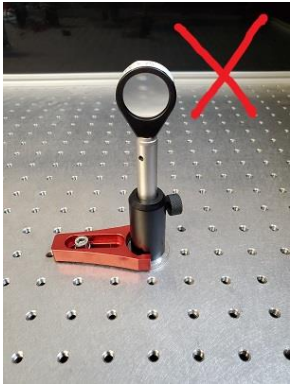  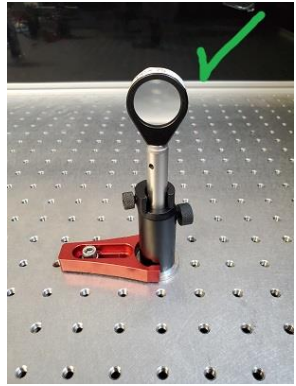  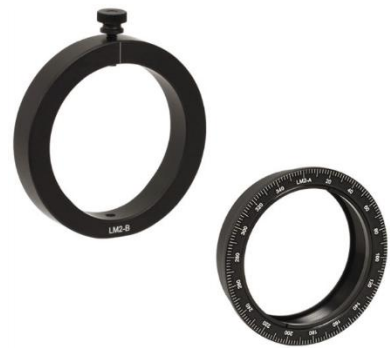

Not easily replaced.           Easily removed/replaced.      or      Nested mount system.

The goal of this skill is that the center of the beam path after the lens is *exactly* the same both with and without the lens in place.

- Remove the lens.
- Adjust the viewing screen such that the exact center of the beam without the lens is indicated
    - if the viewing screen has a target pattern on it then adjust the horizontal and vertical location of the screen such that the beam without the lens is incident upon the center of the target pattern
    - if the viewing screen is simply a piece of paper, or a business card, taped to a post, then use a pencil to mark the exact center of the laser beam on the paper/card

- Replace the lens in the system. This should be easily done if a post clamp is used, or the lens is mounted in a nested mount. If a post clamp is used, just put the post back into the post holder and the clamp should set the height of the lens.
- Observe if the enlarged beam is still *exactly* centered on the same location (center of the target pattern, or the pencil mark on the blank screen)

### Evaluation Checklist

|   | |
|---|---|
|   | **Laser terminates on beam block, or viewing screen after the lens**<br>• the laser always safely terminates on a beam block or viewing screen |
|   | **Lens is easily removed and quickly replaced to the same exact location**<br>• If the lens is mounted to a post, then *a post collar is used* to fix the height and the thumbscrew of the post holder is loosened to remove the lens, post, post collar assembly,<br>• or the student used a flipper mount adapter to mount the lens to the post |
|   | **The beam centers with and without the lens are exactly the same** |
|   | **Handling of the lens**<br>• Did they not touch the surface of the lens with their bare fingers (are there any finger prints on the surfaces of the lens) |
|   | **Optical hardware properly secured to the table**<br>• Table clamps are used properly, all bolts on optical hardware have washers |

# Acknowledgements

The author would like to acknowledge Sebastian Hofferberth and the Institut für Angewandte Physik at the Universität Bonn for providing the sabbatical support, workspace and equipment during the time of writing these tutorials, and Anna Katharina Victoria Rosenberg (Vicky) for her feedback, help, and invaluable insight from a student's perspective.